\documentclass[preprint,12pt]{elsarticle}
\usepackage{epsfig} 
\usepackage{graphicx}
\usepackage[ansinew]{inputenc}
\usepackage{epstopdf}
\usepackage{psfrag}
\usepackage{amsmath}
\usepackage{amssymb}
\usepackage{float}
\usepackage{multirow}
\usepackage{lineno}
\usepackage{setspace}
\usepackage{flushend}
\usepackage{array}
\usepackage{styles/tabu}
\usepackage{bm}
\usepackage{color}
\usepackage{url}
\usepackage{natbib}
\usepackage{fixltx2e}
\newcolumntype{L}[1]{>{\raggedright\let\newline\\\arraybackslash\hspace{0pt}}m{#1}}
\newcolumntype{C}[1]{>{\centering\let\newline\\\arraybackslash\hspace{0pt}}m{#1}}
\newcolumntype{R}[1]{>{\raggedleft\let\newline\\\arraybackslash\hspace{0pt}}m{#1}}
\usepackage[figuresright]{rotating}
\usepackage{subfigure}
\usepackage[title,titletoc,toc]{appendix}
\usepackage{mathrsfs}
\usepackage{amsfonts}
\usepackage{svgcolor}
\usepackage{styles/ieeetrantools}
\usepackage{makecell}

\usepackage{xcolor}
\definecolor{verde}{rgb}{0.0, 0.5, 0.0}
\definecolor{rojo}{rgb}{0.7, 0, 0.0}
\definecolor{grisClaro}{rgb}{0.65, 0.65, 0.65}

\def \tinyREF   {{\scriptscriptstyle \! r\!e\!f}}
\def \tinyREFVAP   {{\scriptscriptstyle \! r\!e\!f_{\!v}}}

\def \tinyREFTWO   {{\scriptscriptstyle \! r\!e\!f_{\!2}}}
\def \tinySEC   {{\scriptscriptstyle \! s\!e\!c}}
\def \tinyPCM   {{\scriptscriptstyle \! p\!c\!m}}
\def \tinyEFF   {{\scriptscriptstyle \! e\!f\!f}}
\def \tinyHOT   {{\scriptscriptstyle \! h\!o\!t}}
\def \tinyCOLD   {{\scriptscriptstyle \! c\!o\!l\!d}}
\def \tinyPCMSOL   {{\scriptscriptstyle \! p\!c\!m_{\!s}}}
\def \tinyPCMLIQ   {{\scriptscriptstyle \! p\!c\!m_{\!l}}}

\def \tinyMAX   {{\scriptscriptstyle \! m\!a\!x}}
\def \tinyMIN   {{\scriptscriptstyle \! m\!i\!n}}

\def \tinyINT    {{\scriptscriptstyle \!i\!n\!t}}
\def \tinyEXT    {{\scriptscriptstyle \!e\!x\!t}}
\def \tinyIN    {{\scriptscriptstyle i\!n}}
\def \tinyOUT    {{\scriptscriptstyle o\!u\!t}}
\def \tinyENV    {{\scriptscriptstyle \!e\!n\!v}}
\def \tinyCOND    {{\scriptscriptstyle c\!o\!n\!d}}
\def \tinyCONV    {{\scriptscriptstyle c\!o\!n\!v}}
\def \tinyCONVINT    {{\scriptscriptstyle c\!o\!n\!v\!,i\!n\!t}}

\def \tinyCONDINT    {{\scriptscriptstyle c\!o\!n\!d\!,i\!n\!t}}
\def \tinyCONDWALL    {{\scriptscriptstyle c\!o\!n\!d\!,w\!a\!l\!l}}

\def \tinyCONVEXT    {{\scriptscriptstyle c\!o\!n\!v\!,e\!x\!t}}

\def \tinyWALL   {{\scriptscriptstyle w\!a\!l\!l}}
\def \tinyLAT   {{\scriptscriptstyle l\!a\!t}}
\def \tinyLATMIN    {{\scriptscriptstyle \!l\!a\!t\!-}}
\def \tinyLATMAX    {{\scriptscriptstyle \!l\!a\!t\!+}}
\def \tinyINTLATMIN    {{\scriptscriptstyle \!i\!n\!t\!-}}
\def \tinyINTLATMINEXTENSO    {{\scriptscriptstyle \!i\!n\!t\!,l\!a\!t\!-}}
\def \tinyTES  {{\scriptscriptstyle \!T\!E\!S}}

\def \tinyRAT  {{\scriptscriptstyle r\!a\!t}}
\def \tinyCHARGE   {{\scriptscriptstyle \! c\!h\!a\!r}}
\def \tinyDCHARGE   {{\scriptscriptstyle \! d\!c\!h\!a\!r}}

\def \tinyLAY {{\scriptscriptstyle \! l\!a\!y}}
\def \tinyNLAY {{n_\tinyLAY}}
\def \tinyPCMOne   {{\scriptscriptstyle \! p\!c\!m\!,1}}
\def \tinyPCMTwo   {{\scriptscriptstyle \! p\!c\!m\!,2}}
\def \tinyPCMK   {{\scriptscriptstyle \! p\!c\!m\!,k}}
\def \tinyPCMKMenos {{\scriptscriptstyle \! p\!c\!m\!,k-1}}
\def \tinyPCMKMas {{\scriptscriptstyle \! p\!c\!m\!,k+1}}
\def \tinyPCMNLAY   {{\scriptscriptstyle \! p\!c\!m\!,\tinyNLAY}}

\def \Nu {{\mathsf{N\hspace{-0.03em}u}}}
\def \Re {{\mathsf{R\hspace{-0.03em}e}}}
\def \Ra {{\mathsf{R\hspace{-0.03em}a}}}
\def \Gr {{\mathsf{G\hspace{-0.03em}r}}}
\def \Pr {{\mathsf{P\hspace{-0.08em}r}}}
\def \Gz {{\mathsf{G\hspace{-0.03em}z}}}
\def \NTU {{\mathsf{N\hspace{-0.04em}T\hspace{-0.04em}U}}}

\journal{Applied Thermal Engineering}

\begin{document}
	
\bstctlcite{bibliografia:BSTcontrol}

\begin{frontmatter}

\title{Novel scheme for a PCM-based cold energy storage system. Design, modelling, and simulation\footnote{© 2018. This manuscript version is made available under the CC-BY-NC-ND 4.0 license \url{https://creativecommons.org/licenses/by-nc-nd/4.0/}. The link to the formal publication is \url{https://doi.org/10.1016/j.applthermaleng.2017.12.088}}}

\author{
    Guillermo Bejarano, 
    Jos{\'e} J. Suffo, 
    Manuel Vargas, 
    Manuel G. Ortega \\
    Department of Systems Engineering and Automatic Control, University of Seville, Spain \\
    \{gbejarano, jossufagu, mvargas, mortega\}@us.es 
		}

\begin{abstract}

{\emph{This paper studies the design and dynamic modelling of a novel thermal energy storage (TES) system combined with a refrigeration system based on phase change materials (PCM). Cold-energy production supported by TES systems is a very appealing field of research, since it allows flexible cold-energy management, combining demand fulfilment with cost reduction strategies. The paper proposes and compares two different simulation models for a cold-energy storage system based on PCM. First, a continuous model is developed, the application of which is limited to decoupled charging/discharging operations. Given such conditions, it is a relatively precise model, useful for the tuning of the TES parameters. The second proposed model is a discrete one, which, despite implementing a discrete approximation of the system behaviour, allows to study more general conditions, such as series of partial charging/discharging operations. Simulation results of both models are compared regarding decoupled charging/discharging operations, and the ability of the discrete model to represent more realistic partial operations is analysed.}}

\end{abstract}


\end{frontmatter}


\section{Introduction} \label{Introduction}

Nowadays, a great deal of energy is required by refrigeration systems, which impact global energy and economical balances. This demand is growing both in industrial or commercial sectors as well as for domestic use \cite{Rasmussen2005}. According to some specialised reports, up to 30\% of the total consumed energy around the world is linked to Heating, Ventilating, and Air Conditioning (HVAC) systems \cite{buzelin2005experimental,jahangeer2011numerical}, while the latest Residential Energy Consumption Survey (RECS) quantifies that around 28\% of the overall consumed energy in the United States is tied to refrigerators and HVAC systems \cite{RECS_US_2009}.

Recently, some research effort is being focused, not only on cold-energy production, but also on cold-energy management. A very interesting strategy, in this context, involves the so-called thermal energy storage (TES) systems, where a thermal reservoir is used to store the excess of cold-energy produced and to release it when necessary. One of the main advantages of these storage systems is the possibility of relaxing the operating conditions, enabling the design of smaller setups and making a more efficient use of them, along with a significant reduction in energy consumption \cite{dincer2002bthermal}. In fact, the availability of auxiliary elements, able to store the energy surplus, will possibly avoid the need of oversizing the system in the design stage, in order to face peak-demand periods
\cite{maccracken2004thermal}.

In the context of refrigeration systems, the opportunity of scheduling, in order to take advantage of available energy low-price time slots, where global energy demand is lower, is very appealing from an economic point of view. Nevertheless, this must be compatible with the need of satisfying the cold-energy demand in real time. Here comes up the idea of counting on a cold-energy reservoir, to drain energy out from in peak-demand periods, avoiding the need of cold-energy generation in such highly-priced periods (\emph{peak-shifting}) \cite{dincer2002bthermal,rismanchi2012energy}.

Phase change materials (PCM) represent a very interesting choice when it comes to devising a TES system, compared to sensible-heat materials, due to their convenient thermodynamic properties for heat transfer, featuring not only a suitable heat capacity, but also a relatively broad latent zone, where the temperature remains almost constant despite energy variations \cite{mehling2008heat}. This enables the storage of the same amount of energy in a lower volume, along with more efficient heat transfer, since the temperature difference between the cold and the hot sources remains steady through the heat exchanger length. Or{\'o} \emph{et al.} provide a classification of actual solid-liquid phase change materials designed for cold-thermal energy storage applications, enumerating all relevant thermodynamic properties, such as melting temperature and latent heat \cite{oro2012review}.

Clearly, design is key for these kind of systems, from choosing the right phase change material, to the proper scaling of the system for the intended application, etc. Besides system design, however, system management is also decisive, in order to optimise some performance indices. Regarding both these aspects, some interesting works by Wang \emph{et al.} can be pointed out \cite{wang2007anovel,wang2007bnovel,wang2007cnovel}. In the first one, the authors undertake the design of a large-scale HVAC plant, backed up by a ring of PCM-based TES tanks \cite{wang2007anovel}. The second one is focused on modelling this complex system \cite{wang2007bnovel}, while the last one suggests a suitable control policy for activation/deactivation of the individual TES elements, in order to attain satisfactory performance levels \cite{wang2007cnovel}. Mosaffa \emph{et al.}, on the other hand, rely on exergetic analysis as a means for quantitatively assessing the performance of HVAC systems where PCM-based TES modules are integrated \cite{mosaffa2014advanced,mosaffa2013thermal}. Again, the proposed control strategy makes a combined usage of the different modules. The model predictive control framework (MPC) has also been successfully applied to PCM-backed refrigeration facilities in a number of works with different goals, such as minimizing deviations in electric energy consumption \cite{shafiei2014model}, guaranteeing product quality in long-term storage subject to minimization of energy consumption \cite{schalbart2015ice}, or prediction of thermal energy demand in refrigerated freight transport \cite{shafiei2015model}, among others.

Many theoretical works about the mathematical modelling of latent heat TES have been proposed over the years. Some very complete reviews, for instance those by Verma and Singal \cite{verma2008review}, Or{\'o} \emph{et al.} \cite{oro2012review}, and Dutil \emph{et al.} \cite{dutil2011review} include different modelling paradigms, according to the selected geometry. In particular, packed beds are one of the most successful technologies, since thermal conductivity of most PCMs is low, thus increasing the surface/volume ratio is interesting to improve heat transfer. A packed bed consists of a volume including a large number of PCM capsules. During the heat charging process, warm heat transfer fluid flows through the volume while melting the PCM in the capsules. Conversely, during the heat discharging process, cold heat transfer fluid flows through the system solidifying the PCM in the capsules. In these processes, the PCM absorbs or releases a large amount of energy as latent heat. Saitoh and Hirose put in evidence this idea both theoretically and experimentally \cite{saitoh1986high}; however, the heat transfer increase provided by this technology comes at the price of a significant pressure drop and a higher initial cost.

A general model is provided by Zhang \emph{et al.}, where thermal performance of both melting and solidification processes is analysed \cite{zhang2001general}. The instantaneous temperature distribution, the instantaneous heat transfer rate, and the thermal storage capacity can be computed. Benmansour \emph{et al.} developed a two-dimensional transient numerical model of a cylindrical packed bed randomly filled with uniformly sized spheres using air as a working fluid \cite{benmansour2006experimental}. Furthermore, B{\'e}d{\'e}carrats \emph{et al.} have addressed the model of a cylindrical tank filled with encapsulated PCM while studying the impact of supercooling \cite{bedecarrats2009astudy,bedecarrats2009bstudy}. They state that the optimum charging mode is obtained in the case of the vertical position, since the motion due to natural convection is in the same direction as the forced convection. Cheralathan \emph{et al.} conclude that lower inlet heat transfer fluid (HTF) temperature reduces the supercooling of the PCM and the total charging time \cite{cheralathan2007effect}. However, the reduction of heat-transfer fluid temperature worsens the performance of the refrigeration system. Moreover, low porosity improves storage capacity of the system, but the charging/discharging rate is reduced. Ismail and Henriquez propose a simplified one-dimensional transient model of the PCM capsule stocked in a cylindrical tank \cite{ismail2002numerical}. The convection present in the liquid phase of the PCM is treated by using an effective heat conduction coefficient. The solution of the differential equations is realised by the finite-difference approximation and a moving mesh inside the spherical capsule. Furthermore, MacPhee and Dincer study the melting and freezing of water in spherical capsules using ANSYS, GAMBIT, and FLUENT 6.0 \cite{macphee2009thermodynamic}. The models are validated using experimental data. Energy and exergetic efficiency have been also analysed. Higher temperature difference between the HTF and PCM shows to be most optimal exergetically, but least optimal energetically due to entropy generation. Higher temperature difference also increases the rates of charging/discharging. A simple analytical model is developed by the same team to reproduce the results of the numerical models \cite{macphee2009thermal}.

Nevertheless, in the packed bed technology, the same HTF is used to charge and discharge the system. In our application two different fluids might be used: the refrigerant in the charging process and other HTF in the discharging process, since the cold-thermal energy storage tank is engineered to complement an existing vapour-compression refrigeration facility. Therefore, a novel hybrid structure is proposed, where the tank is filled with PCM spheres, bathed in the so-called \emph{intermediate fluid}, which is a fluid with high thermal conductivity and low heat capacity. In addition to the PCM capsules, two bundles of pipes, corresponding to the refrigerant and secondary fluid, are also in contact with the intermediate fluid and running through the tank. This setup is different from others found in similar applications; for instance, Tay \emph{et al.} \cite{tay2012experimental}, directly use bulk PCM inside the tank, instead of encapsulated PCM. PCM-based TES systems found in facilities from CRISTOPIA Energy Systems do use encapsulated PCM, however, they do not use dual bundle of pipes, rather they use a common pipe, where the same fluid flows, acting as cold source when charging or as hot source when discharging. The bathing fluid is based on the proprietary STL technology \cite{STL_CRISTOPIA}.

In this paper, the system modelling strongly relies on a totally parameterisable approach, which allows to adjust the PCM thermodynamic properties, encapsulation features, number and size of PCM capsules, among many other factors, in order to meet the operation requirements. Two different models are proposed: firstly, a simplified continuous model is proposed, which comprises only a limited description of the system dynamics. Secondly, given the limitations of the continuous model, a new discrete model in which the PCM capsules are conceptually divided in a given set of spherical layers is proposed, able to represent the system evolution during any series of partial charging/discharging operations.

The paper is organised as follows. Section \ref{secSystemDescription} describes the refrigeration plant which the TES system is being added to and the way the interconnection is being projected, including the particular constructive details, specifications, and parameters, so that the interested readers have full information about the proposed setup. Section \ref{secSystemModelling} details the underlying modelling of the proposed cold-thermal energy storage system. Section \ref{secCompleteDiscreteModel} points out the main limitations of the previous model and the interest of a new discrete spatial model, which provides computational feasibility of certain practical operational modes. This same section provides several comparative simulations of full charge/discharge or sequences of partial charging/discharging operations. Section \ref{secConclusions} summarises the main conclusions of the study. Eventually, \ref{appendixCorrelations} goes into detail on the particular correlations used, while \ref{appendixEquationSummary} summarises the nonlinear equations which define the continuous model.


\section{System description} \label{secSystemDescription}

Tables \ref{tabSymbols} and \ref{tabSubscriptsSuperscripts} describe the list of symbols and the subscript/superscript notation followed along the modelling description of the thermal storage system. Fluids properties, such as $c_{\!p}$, $\beta$, $\kappa$, $\mu$, and $\rho$ are not constant, but dependent on temperatures, pressures, etc. Their instantaneous values are rigorously considered in the simulation models, provided by the \emph{CoolProp} tool \cite{CoolProp}. Nonetheless, for the sake of clarity, in the following description, time dependence of these parameters is not made explicit.

\begin{table}[H]
	\centering
	\caption{List of symbols}
	\label{tabSymbols}
	\scalebox{0.6}[0.6]{ \tabulinesep=0.5mm
		\begin{tabu} { L{1.3cm} L{7.7cm} L{1.8cm} | L{1.3cm} L{6.5cm} L{1.9cm} }
			\multicolumn{3}{c|}{\emph{\bf Latin symbols}} & \multicolumn{3}{c}{\emph{\bf Greek symbols}} \\ 
			\Xhline{3\arrayrulewidth}
			\emph{Symbol} & \emph{Description} & \emph{Units} & \emph{Symbol} & \emph{Description} & \emph{Units} \\ 
			\Xhline{3\arrayrulewidth}
			$C$ & Heat capacity rate & W K\textsuperscript{-1} & $\alpha$ & Convective heat transfer coefficient & W m\textsuperscript{-2} K\textsuperscript{-1} \\
			\Xhline{2\arrayrulewidth}
			$c_{\!p}$ & Specific heat at constant pressure & J kg\textsuperscript{-1} K\textsuperscript{-1} & $\beta$ & Coefficient of thermal expansion & K\textsuperscript{-1} \\ 
			\Xhline{2\arrayrulewidth}
			$cd$ & Binary value defining charging/discharging process & & $\gamma$ & \emph{Charge ratio} & p.u. \\ 
			\Xhline{2\arrayrulewidth}
			$e$ & Thickness & m & $\varepsilon$ & Effectiveness & p.u. \\
			\Xhline{2\arrayrulewidth}
			$f$ & Generic function & & $\zeta$ & Fraction of a given length, $l$ & p.u. \\
			\Xhline{2\arrayrulewidth}
			$G$ & Mass flux (rate of mass flow per unit area) & kg m\textsuperscript{-2} s\textsuperscript{-1} & $\kappa$ & Thermal conductivity & W m\textsuperscript{-1} K\textsuperscript{-1} \\
			\Xhline{2\arrayrulewidth}
			$g$ & Acceleration of gravity & m s\textsuperscript{-2} & $\mu$ & Dynamic viscosity & kg m\textsuperscript{-1} s\textsuperscript{-1} \\
			\Xhline{2\arrayrulewidth}
			$h$ & Specific enthalpy & J kg\textsuperscript{-1} & $\rho$ & Density & kg m\textsuperscript{-3} \\
			\Xhline{2\arrayrulewidth}
			$l$ & Individual pipe length & m & $\sigma$ & Surface tension & N m\textsuperscript{-1} \\
			\Xhline{2\arrayrulewidth}
			$m$ & Mass & kg & $\phi$ & Auxiliary function used in the Klimenko's method & \\
			\Xhline{2\arrayrulewidth}
			$\dot m$ & Mass flow rate & kg s\textsuperscript{-1} & $\chi$ & Vapour quality & p.u. \\
			\Xhline{2\arrayrulewidth}
			$n$ & Number of elements & & $\psi$ & Friction factor & p.u. \\
			\Xhline{2\arrayrulewidth}
			$P$ & Pressure & Pa & & & \\
			\Xhline{2\arrayrulewidth}
			$\dot Q$ & Transferred cooling power & W & & & \\
			\Xhline{2\arrayrulewidth}
			$q$ & Internal heat flux & W m\textsuperscript{-2} & \multicolumn{3}{c}{\emph{\bf Heat transfer dimensionless numbers}} \\
			\Xhline{2\arrayrulewidth}
			$R$ & Thermal resistance & K W\textsuperscript{-1} & $\Gr$ & \emph{Grashof} number & \\
			\Xhline{2\arrayrulewidth}
			$r$ & Internal radius & m & $\Gz$ & \emph{Graetz} number & \\
			\Xhline{2\arrayrulewidth}
			$T$ & Temperature & ºC & $\Nu$ & \emph{Nusselt} number & \\
			\Xhline{2\arrayrulewidth}
			$t$ & Time & s & $\Pr$ & \emph{Prandtl} number & \\
			\Xhline{2\arrayrulewidth}
			$U$ & Internal energy & J & $\Ra$ & \emph{Rayleigh} number & \\
			\Xhline{2\arrayrulewidth}
			$V$ & Volume & m\textsuperscript{3} & $\Re$ & \emph{Reynolds} number & \\
			\Xhline{2\arrayrulewidth}
			$\boldsymbol{x}$ & State vector & & & & \\
			\Xhline{2\arrayrulewidth}
	\end{tabu}}
\end{table}

\begin{table}[H]
	\centering
	\caption{Subscript/superscript notation}
	\label{tabSubscriptsSuperscripts}
	\scalebox{0.65}[0.65]{\tabulinesep=0.5mm
		\begin{tabu} { L{1.3cm}  L{6cm} | L{1.3cm}  L{6cm} }
			\multicolumn{2}{c|}{\emph{\bf Subscripts}} & \multicolumn{2}{c}{\emph{\bf Superscripts}} \\ 
			\Xhline{3\arrayrulewidth}
			\emph{Symbol} & \emph{Description} & \emph{Symbol} & \emph{Description}\\ 
			\Xhline{3\arrayrulewidth}
			$\tinyCOLD$ & Cold source & $\tinyCHARGE$ & Charging process \\ 
			\Xhline{2\arrayrulewidth}
			$\tinyEFF$ & \emph{Effective} & $\tinyCOND$ & Conduction \\
			\Xhline{2\arrayrulewidth} 
			$\tinyENV$ & Environment & $\tinyCONV$ & Convection \\ 
			\Xhline{2\arrayrulewidth}
			$\tinyHOT$ & Hot source & $\tinyDCHARGE$ & Discharging process \\
			\Xhline{2\arrayrulewidth}
			$\tinyINT$ & Intermediate fluid & $\tinyEXT$ & External \\ 
			\Xhline{2\arrayrulewidth}
			$\tinyLAY$ & Layer of the PCM capsule & $\tinyIN$ & Input/Inlet \\ 
			\Xhline{2\arrayrulewidth}
			$\tinyPCM$ & PCM & $\tinyINT$ & Internal \\  
			\Xhline{2\arrayrulewidth}
			$\tinyPCMLIQ$ & PCM in liquid phase & $\tinyINTLATMIN$ & Abbreviation of $\tinyINTLATMINEXTENSO$ \\ 
			\Xhline{2\arrayrulewidth}
			$\tinyPCMSOL$ & PCM in solid phase & $\tinyLAT$ & Latent state \\ 
			\Xhline{2\arrayrulewidth}
			$\tinyREF$ & Refrigerant & $\tinyLATMAX$ & Maximum latent enthalpy \\ 
			\Xhline{2\arrayrulewidth}
			$\tinyREFTWO$ & Refrigerant as two-phase fluid & $\tinyLATMIN$ & Minimum latent enthalpy \\ 
			\Xhline{2\arrayrulewidth}
			$\tinyREFVAP$ & Refrigerant as superheated vapour & $\tinyMAX$ & Maximum \\ 
			\Xhline{2\arrayrulewidth}
			$\tinySEC$ & Secondary fluid & $\tinyMIN$ & Minimum \\ 
			\Xhline{2\arrayrulewidth}
			$\tinyTES$ & Thermal Energy Storage & $\tinyOUT$ & Output/Outlet \\ 
			\Xhline{2\arrayrulewidth}
			 & & $\tinyRAT$ & Ratio \\ 
			\Xhline{2\arrayrulewidth} 
			 & & $\tinyWALL$ & Wall \\ 
		    \Xhline{2\arrayrulewidth} 
	\end{tabu}}
\end{table}

\subsection{TES on the actual mother refrigeration plant}

\begin{figure}[htbp]
    \centerline{\includegraphics[width=12cm,trim = 0 55 0 90,clip]
    	{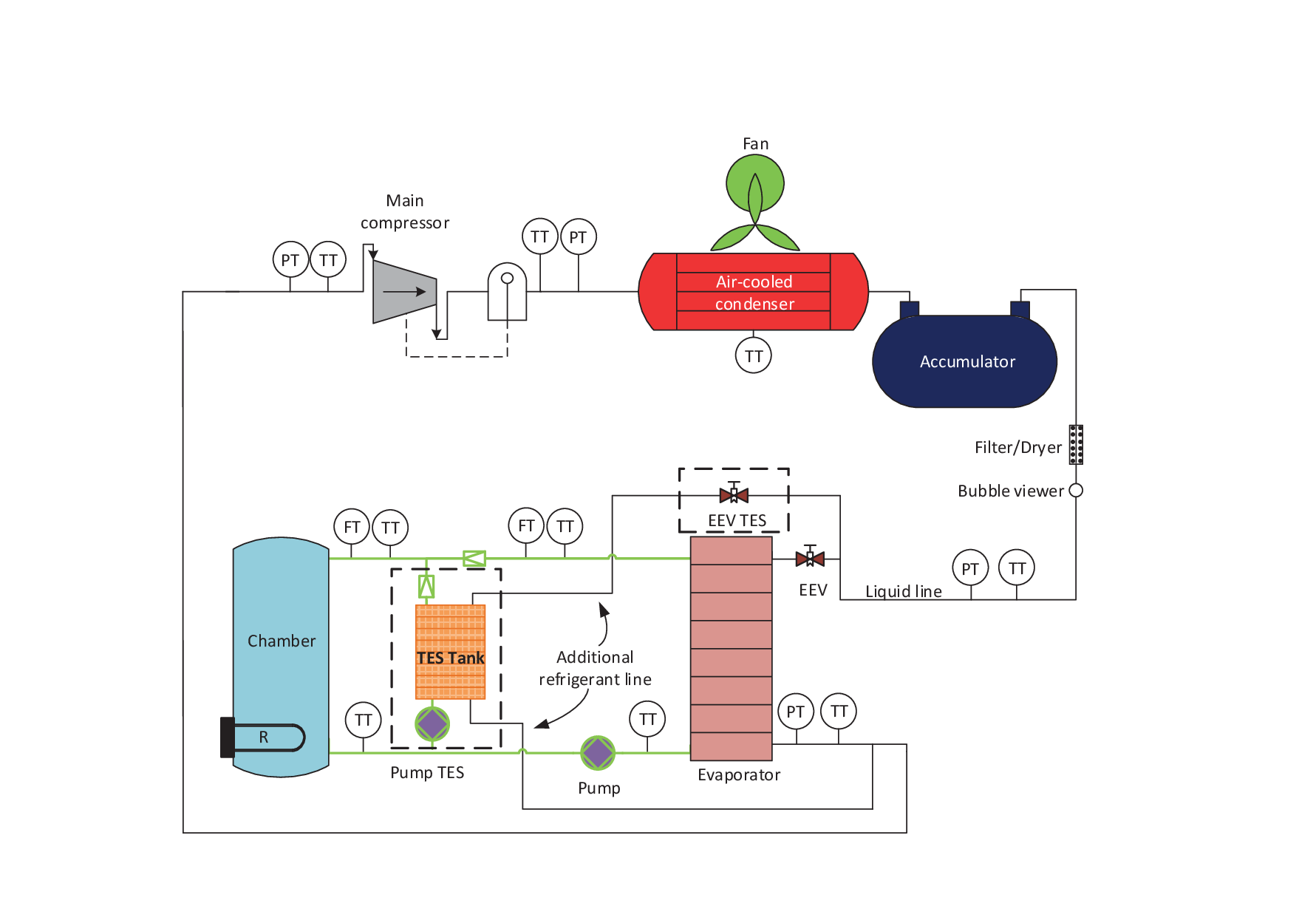}}
    \caption{Schematic diagram of the experimental refrigeration plant.}
    \label{figEsquemaPlanta}
\end{figure}

Figure \ref{figEsquemaPlanta} shows a diagram of a part of the refrigeration experimental facility located at the Department of Systems Engineering and Automation of University of Seville, in which a TES tank is projected to be included (marked with a dash line in diagram).

The cycle works with R404a as refrigerant. The evaporator is a brazed-plate heat exchanger, being its secondary fluid an aqueous solution of around {60\%} propylene glycol. The secondary mass flow is regulated by a liquid pump, which is controlled by a variable frequency drive, like the compressor. The secondary inlet temperature can be regulated using the electric resistance located at the chamber. It is intended that the temperature level at the evaporator should satisfy cooling demand at $-20^{\circ}\mathrm{C}$; this temperature has been selected keeping in mind typical reference levels for refrigeration systems. The condenser is an air-cooled cross-flow tube heat exchanger. The electronic expansion valve is controlled using Pulse-Width Modulation (PWM). Regarding instrumentation, thermocouples and pressure transducers have been placed at selected points to compute all refrigerant enthalpies using thermodynamic functions, while a magnetic-inductive flow sensor is available to measure the volumetric flow rate of the glycol-based solution. A detailed description of this plant can be found in the related literature \cite{GB_JE_2015,bejarano2015multivariable}.

A TES tank is intended to be added to this refrigeration plant, in parallel with the refrigerated chamber working at $-20^{\circ}\mathrm{C}$. An additional refrigerant line needs to be deployed, together with its respective expansion valve, enabling the TES charging operation. The discharging operation, on the other hand, will be possible through a new pump enabling the recirculation of the secondary fluid from the chamber to be cooled, labelled as \emph{Pump TES} in Figure \ref{figEsquemaPlanta}.

\subsection{A novel layout proposal of the TES system}

To the best of the authors' knowledge, the proposed TES system setup is a novel one, clearly different from existing designs that can be found in similar applications, as reported in Section \ref{Introduction}. The design was highly constrained by the configuration and operating conditions imposed by the existing refrigeration plant. According to this, while a single fluid conduct going through the PCM tank is the most common, our design needed to make use of a refrigerant fluid as cold source and the secondary fluid as HTF to cool the refrigerated chamber. Apart from that, the available refrigerant and secondary fluid inlet temperatures, pressures, and mass flow rates, on the one hand, and the aimed charging and discharging time intervals (3-4 hours were regarded as most desirable), on the other hand, were the main factors dictating the workable layouts. After several design iterations, the new design was conceived.

\begin{figure}[h]
	\centerline{\includegraphics[width=7cm,trim = 0 105 0 0,clip]
	{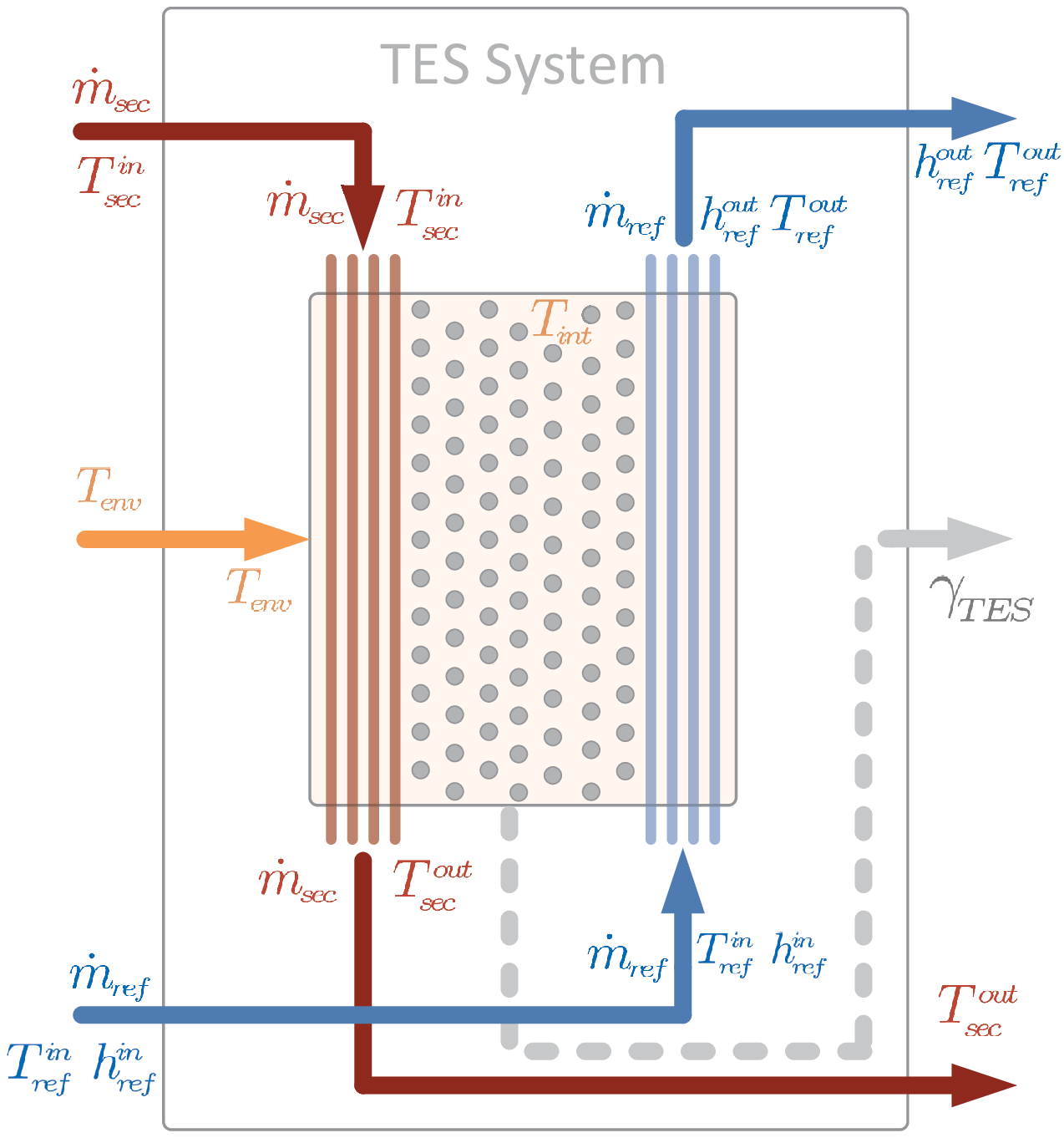}}
	\caption{Schematic picture of the proposed configuration of the TES tank and input-output conceptualisation of the TES system.}
	\label{figTEStank}
\end{figure}

A particular efficient phase change material has to be chosen, as well as its appearance or presentation. In this design, the TES tank is being filled with macro-encapsulated PCM, where the raw material is enclosed in spherical polymer capsules, whose diameter is typically above 1 cm. The PCM spheres are bathed in the so-called intermediate fluid, which is a fluid with high thermal conductivity and low heat capacity. In addition to the PCM capsules, two bundles of pipes, corresponding to the refrigerant and secondary fluid, are also in contact with the intermediate fluid and running through the tank. This way, when the refrigerant is flowing, the tank acts as an evaporator, where the refrigerant evaporates while extracting heat from the intermediate fluid (and eventually from the PCM capsules). On the other hand, the secondary fluid is a glycol-water solution, which, when flowing, transfers heat to the intermediate fluid and then to the PCM spheres. Figure \ref{figTEStank} shows a schematic picture of the proposed setup for the TES tank. It should be noticed that this is just a conceptual scheme. Of course, from a constructive point of view, all pipe lines must be distributed in such a way that homogeneous heat transfer is achieved all around the TES tank.

Conceptual input-output representation of the system can be also noticed in Figure \ref{figTEStank}. The inputs are the refrigerant mass flow rate, $\dot m_\tinyREF(t)$, and secondary fluid mass flow rate, $\dot m_\tinySEC(t)$, essentially used as control inputs during the charging or discharging cycle, respectively. Other input signals are the inlet temperature of the secondary fluid, $T_\tinySEC^\tinyIN(t)$, the inlet temperature and the specific enthalpy of the refrigerant, $T_\tinyREF^\tinyIN(t)$ and $h_\tinyREF^\tinyIN(t)$, and the ambient temperature outside the TES tank, $T_\tinyENV(t)$. Clearly, the refrigerant pressure, intermediate fluid pressure, and secondary fluid pressure also affect the system, but they are not shown in Figure \ref{figTEStank}, since they are assumed to be constant, according to the equilibrium conditions the whole refrigeration plant is operating at. Note that if the refrigerant pressure varies along time, according to the behaviour of the refrigeration cycle, the thermodynamic conditions of the refrigerant flowing into the TES tank will vary too, in such a way that the system will adapt to the new refrigerant inlet conditions. The values of the variables mentioned at the nominal operating point of the experimental plant are shown in Table \ref{tabNominalOperatingConditions}. 

\begin{table}[h]
    \centering
    \caption{Nominal operating conditions}
    \label{tabNominalOperatingConditions}
    \scalebox{0.7}[0.7]{ \tabulinesep=0.5mm{}
    \begin{tabu} {L{1.5cm}  L{11cm} L{1.5cm} L{2cm}}
   		\emph{Symbol} & \emph{Description} & \emph{Value} & \emph{Units} \\ 
		\Xhline{5\arrayrulewidth}
   		$\dot m_\tinyREF$ & Refrigerant mass flow rate & 0.00918 & kg 	s\textsuperscript{-1} \\ 
   		\Xhline{2\arrayrulewidth}
   		$h_\tinyREF^\tinyIN$ & Specific enthalpy of the refrigerant at the TES tank inlet & 255000 & J kg\textsuperscript{-1} K\textsuperscript{-1} \\ 
   		\Xhline{2\arrayrulewidth}
   		$P_\tinyREF$ & Refrigerant pressure & 126500 & Pa \\ 
   		\Xhline{2\arrayrulewidth}
   		$T_\tinyREF^\tinyIN$ & Temperature of the refrigerant at the TES tank inlet & -41.08 & ºC \\ 
   		\Xhline{3\arrayrulewidth}
   		$\dot m_\tinySEC$ & Secondary mass flow rate & 0.074 & kg s\textsuperscript{-1} \\ 
   		\Xhline{2\arrayrulewidth}
   		$P_\tinySEC$ & Secondary fluid pressure & 100000 & Pa \\ 
   		\Xhline{2\arrayrulewidth}
   		$T_\tinySEC^\tinyIN$ & Temperature of the secondary fluid at the TES tank inlet & -20 & ºC \\ 
   		\Xhline{2\arrayrulewidth}
	\end{tabu}}
\end{table}

Regarding the output signals shown in Figure \ref{figTEStank}, apart from the outlet thermodynamic conditions of both the refrigerant and the secondary fluid,  $\gamma_\tinyTES(t)$ (indicated by discontinuous line since it is a state variable) is defined as the cold-thermal energy storage ratio of the TES system, to be referred as the \emph{charge ratio} for short, from now on. Its meaning will be discussed in the following paragraphs.

\begin{figure}[htbp]
    \centerline{\includegraphics[width=14cm,trim = 0 18 0 20,clip]
    {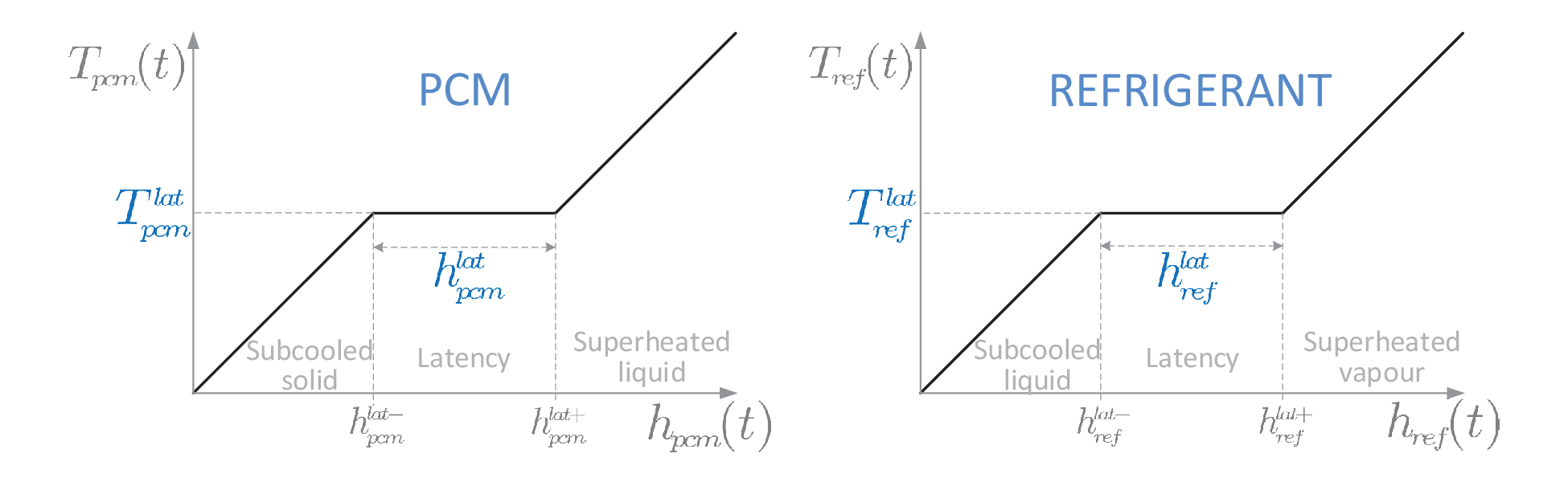}}
    \caption{Temperature-enthalpy diagrams of PCM and refrigerant.}
    \label{figThDiagrams}
\end{figure}

During normal operation of our system, both materials, the refrigerant and the PCM, may experience phase change. In particular, the PCM may be in solid phase, liquid phase, or in transition between both. On the other hand, the refrigerant may appear in liquid phase, vapour phase, or two-phase. Figure \ref{figThDiagrams} shows the temperature-enthalpy diagrams of the PCM and the refrigerant, in the latter case at constant pressure since the refrigerant pressure drop is expected to be negligible along the TES tank. Note that, in the case of the PCM, $h_\tinyPCM^\tinyLATMIN$ only represents a reference level on the PCM specific enthalpy $h_\tinyPCM$, in such a way that $h_\tinyPCM^\tinyLATMIN$, $h_\tinyPCM^\tinyLATMAX$, and $h_\tinyPCM^\tinyLAT$ are linked by the expression indicated in Equation \eqref{eq:hlat}. 

\begin{equation}
	h_\tinyPCM^\tinyLATMAX \,=\, h_\tinyPCM^\tinyLATMIN \,+\, h_\tinyPCM^\tinyLAT \\
	\label{eq:hlat}
\end{equation}

Concerning $\gamma_\tinyTES(t)$, it represents the instantaneous charging state of the TES system as a fraction of the maximum ''efficient'' cold-energy storage the system is able to take on. More specifically, during the charging and discharging operations, ideally, the PCM should permanently remain in latent state, or, at least, roughly keeping its specific enthalpy inside the interval: $h_\tinyPCM^\tinyLATMIN < h_\tinyPCM(t) < h_\tinyPCM^\tinyLATMAX$. These are the desirable conditions, since heat transfer between the PCM and the intermediate fluid is more efficient, due to its higher capacity and more uniform temperature. According to this, our convention is defining the \emph{maximum ''efficient'' cold-energy storage} in the system, as the energy stored inside the PCM capsules, when their whole volume reaches $h_\tinyPCM^\tinyLATMIN$, i.e. the minimum enthalpy within the latent zone. Reducing the system enthalpy beyond that point to store cold-energy, by taking the PCM to become subcooled solid, is not as efficient as the enthalpy reduction during the latent phase.

Since it is interesting to measure the stored \emph{cold-thermal energy}, instead of the mere thermal energy, we define the \emph{charge ratio}, $\gamma_\tinyTES(t)$, as a normalised index between 0 and 1 in efficient storing conditions, as indicated in Equation \eqref{eq:ChargeRatioDefinition}.

\begin{equation}
   \gamma_\tinyTES(t) \,=\, \frac{U_\tinyTES^\tinyMAX - U_\tinyTES(t)}{U_\tinyTES^\tinyMAX - U_\tinyTES^\tinyMIN}
   \label{eq:ChargeRatioDefinition}
\end{equation}

Since the $n_\tinyPCM$ capsules are supposed to be uniformly distributed inside the TES tank and, in accordance to the assumed high thermal conductivity of the intermediate fluid, it is considered that all PCM capsules will evolve in the same way; hence, the stored thermal energy in the full system can be supposed merely proportional to the thermal energy stored inside each PCM capsule, as shown in Equation Set \eqref{eq:TESEnergyDefinition}, where $U_\tinyPCM(t)$, $U_\tinyPCM^\tinyMIN$, and $U_\tinyPCM^\tinyMAX$ are the instantaneous, minimum, and maximum energy level held inside each PCM capsule, to be defined later, for the models to be presented.

\begin{equation}
   \begin{aligned}
   U_\tinyTES(t) \,&=\, n_\tinyPCM\,U_\tinyPCM(t)\; \\
   U_\tinyTES^\tinyMIN \,&=\, n_\tinyPCM\,U_\tinyPCM^\tinyMIN\; \\
   U_\tinyTES^\tinyMAX \,&=\, n_\tinyPCM\,U_\tinyPCM^\tinyMAX \\
   \end{aligned}
   \label{eq:TESEnergyDefinition}
\end{equation}

Therefore, the \emph{charge ratio} $\gamma_\tinyTES(t)$ can be computed as indicated in Equation \eqref{eq:ChargeRatioDefinition2}.

\begin{equation}
   \gamma_\tinyTES(t) \,=\, \frac{U_\tinyPCM^\tinyMAX - U_\tinyPCM(t)}{U_\tinyPCM^\tinyMAX - U_\tinyPCM^\tinyMIN}
   \label{eq:ChargeRatioDefinition2}
\end{equation}

\subsection{Design specifications} \label{subsecDesignSpecifications}

The TES system previously described has been projected following a detailed set of specifications. In this section the chosen fluids, materials, and other design specifications are given.

The PCM capsules are filled with an ad-hoc designed vegetable oil enclosed in spherical polymer capsules, made of high-density polyethylene. The chosen vegetable oil is a phase change material designed taking the \emph{PureTemp-37}$^{\tiny{\textregistered}}$ \cite{PureTemp-37} as a reference, with the same properties except the melting point, which is -30ºC, instead of -37ºC.

The intermediate fluid is a 60\% v/v ethylene glycol aqueous solution with very high thermal conductivity, providing efficient heat transfer between the refrigerant and the PCM capsules during the charging cycle, or between the PCM capsules and the secondary fluid during the discharging cycle. The mixture has low freezing temperature which meets the requirements inside the tank. Further research has been carried out in order to find other alternative fluids with even better features. A possibility is the use of calcium chloride brines, which have higher thermal conductivity or choosing Pekasol 2000$^{\tiny{\textregistered}}$ \cite{Pekasol2000}, an aqueous solution of environmentally safe alkaline earth formate and acetate with better heat transfer features for the considered application. These possibilities have been evaluated by using the \emph{CoolProp} tool \cite{CoolProp} and they are planned to be studied and analysed in further simulations.

As stated previously, the fluid chosen as refrigerant is R404a, which is a fluid commonly used in refrigeration cycles, specifically in medium-low temperature facilities. It is a mixture of R-125 (pentafluoroethane), R-143a (1,1,1-trifluoroethane), and R-134a (1,1,1,2-tetrafluoroethane).

The secondary fluid is a 60\% v/v propylene glycol aqueous solution. The mixture is suitable in this case since it has low freezing temperature, avoiding clogs and fractures in pipes due to solidification.

The pipes are both made of carbon steel, while the PCM properties, design parameters, and convective heat transfer coefficients considered within the simulations are shown in Tables \ref{tabPCMProperties}, \ref{tabDesignParameters}, and \ref{tabHeatTransferCoefficients}, respectively.

\begin{table}[htbp]
    \centering
    \caption{Phase change material properties}
    \label{tabPCMProperties}
    \scalebox{0.6}[0.6]{ \tabulinesep=0.5mm{}
        \begin{tabu} {L{1.5cm} L{12.5cm} L{2cm} L{2cm}}
            \emph{Symbol} & \emph{Description} & \emph{Value} & \emph{Units} \\ \Xhline{3\arrayrulewidth}
            $c_{\!p_{\,\tinyPCMLIQ}}$ & Specific heat of the PCM in liquid phase at constant pressure & 1990 & J kg\textsuperscript{-1} K\textsuperscript{-1} \\ 
            \Xhline{2\arrayrulewidth}
            $c_{\!p_{\,\tinyPCMSOL}}$ & Specific heat of the PCM in solid phase at constant pressure & 1390 & J kg\textsuperscript{-1} K\textsuperscript{-1} \\ 
            \Xhline{2\arrayrulewidth}
            $h_\tinyPCM^\tinyLAT$ & Specific enthalpy of fusion (latent phase) of the PCM & 145000 & J kg\textsuperscript{-1} \\ 
            \Xhline{2\arrayrulewidth}
            $T_\tinyPCM^\tinyLAT$ & Phase change temperature & -30 & ºC\\ 
            \Xhline{2\arrayrulewidth}$\kappa_\tinyPCM^\tinyLATMAX$ &  Thermal conductivity of PCM in liquid phase & 0.15 & W m\textsuperscript{-1} K\textsuperscript{-1} \\ 
            \Xhline{2\arrayrulewidth}
            $\kappa_\tinyPCM^\tinyLATMIN$ &  Thermal conductivity of PCM in solid phase & 0.25 &  W m\textsuperscript{-1} K\textsuperscript{-1} \\ 
            \Xhline{2\arrayrulewidth}
            $\rho_\tinyPCM^\tinyLATMAX$ & Density of the PCM in liquid phase & 880 & kg m\textsuperscript{-3}\\ 
            \Xhline{2\arrayrulewidth}
            $\rho_\tinyPCM^\tinyLATMIN$ & Density of the PCM in solid phase & 970 & kg m\textsuperscript{-3}\\ 
            \Xhline{2\arrayrulewidth}
    \end{tabu}}
\end{table}

\begin{table} [htbp]
    \centering
    \caption{Design parameters}
    \label{tabDesignParameters}
    \scalebox{0.6}[0.6]{ \tabulinesep=0.5mm {}
	\begin{tabu} {L{1.5cm} L{14cm} L{3.5cm} L{2cm}}
   	\emph{Symbol} & \emph{Description} & \emph{Value} & \emph{Units} \\ \Xhline{3\arrayrulewidth} 
   	$c_{\!p_\tinyINT}$ & Specific heat of the intermediate fluid at constant pressure & 2720.2 -- 3159.6 & J kg\textsuperscript{-1} K\textsuperscript{-1} \\ \Xhline{2\arrayrulewidth}
   	$c_{\!p_{\,\tinyREFVAP}}$ & Specific heat of the refrigerant in vapour phase at constant pressure & 803.19 -- 810.69 & J kg\textsuperscript{-1} K\textsuperscript{-1} \\ 
   	\Xhline{2\arrayrulewidth}
   	$c_{\!p_{\,\tinySEC}}$ & Specific heat of the secondary fluid at constant pressure & 3411.4 -- 3567.1 & J kg\textsuperscript{-1} K\textsuperscript{-1} \\ \Xhline{2\arrayrulewidth}
   	$e_\tinyPCM$ & Thickness of PCM capsule polymer coating & 0.0036 & m \\ \Xhline{2\arrayrulewidth}
   	$e_\tinyREF$ & Thickness of the refrigerant pipe wall& 0.0036 & m \\ \Xhline{2\arrayrulewidth}
   	$e_\tinySEC$ & Thickness of the secondary fluid pipe wall& 0.0036 & m \\ \Xhline{2\arrayrulewidth}
   	$g$ & Acceleration of gravity  & 9.81 & m s\textsuperscript{-2} \\ \Xhline{2\arrayrulewidth}
   	$h_\tinyREF^\tinyLAT$ & Specific enthalpy of vaporization (latent phase) of the refrigerant  & 197770 & J kg\textsuperscript{-1} \\ 
   	\Xhline{2\arrayrulewidth}
   	$l_\tinyREF$ & Refrigerant pipe length & 0.8 & m \\ \Xhline{2\arrayrulewidth}
   	$l_\tinySEC$ & Secondary fluid pipe length & 0.8 & m \\ \Xhline{2\arrayrulewidth}
   	$m_\tinyINT$ & Mass of the intermediate fluid & 56.37 & kg \\ \Xhline{2\arrayrulewidth}
   	$n_\tinyPCM$ & Number of PCM capsules & 400 & \\ \Xhline{2\arrayrulewidth}
   	$n_\tinyREF$ & Number of refrigerant pipes & 50 & \\ \Xhline{2\arrayrulewidth}
   	$n_\tinySEC$ & Number of secondary fluid pipes & 50 & \\ \Xhline{2\arrayrulewidth}
   	$r_{\!\tinyPCM}^{\,\tinyMAX}$ & Maximum internal radius of the PCM capsules & 0.0285 & m \\ 
   	\Xhline{2\arrayrulewidth}
   	$r_{\!\tinyPCM}^{\,\tinyMIN}$ & Minimum internal radius of the PCM capsules & 0.02759 & m \\ 
   	\Xhline{2\arrayrulewidth}
   	$r_{\!\tinyREF}$ & Internal radius of the refrigerant pipe& 0.0218 & m \\ \Xhline{2\arrayrulewidth}
   	$r_{\!\tinySEC}$ & Internal radius of the secondary fluid pipe & 0.0218 & m \\ 
   	\Xhline{2\arrayrulewidth}
   	$\beta_\tinyINT$ & Coefficient of thermal expansion of the intermediate fluid & 0.00044 -- 0.00057 & K\textsuperscript{-1} \\ 
   	\Xhline{2\arrayrulewidth}
   	$\kappa_\tinyINT$ & Thermal conductivity of the intermediate fluid & 0.3905 -- 0.4213 & W m\textsuperscript{-1} K\textsuperscript{-1} \\ 
   	\Xhline{2\arrayrulewidth}
   	$\kappa_\tinyPCM^\tinyWALL$ & Thermal conductivity of the PCM-capsule polymer coating & 0.2 & W m\textsuperscript{-1} K\textsuperscript{-1} \\ \Xhline{2\arrayrulewidth}
   	$\kappa_\tinyREFVAP$ & Thermal conductivity of the refrigerant in vapour phase & 0.0087  -- 0.0095 & W m\textsuperscript{-1} K\textsuperscript{-1} \\ \Xhline{2\arrayrulewidth}
   	$\kappa_\tinyREF^\tinyWALL$ & Thermal conductivity of the refrigerant pipe wall & 45 & W m\textsuperscript{-1} K\textsuperscript{-1} \\ 
   	\Xhline{2\arrayrulewidth}
   	$\kappa_\tinySEC$ & Thermal conductivity of the secondary fluid & 0.3624 -- 0.3692 & W m\textsuperscript{-1} K\textsuperscript{-1} \\ \Xhline{2\arrayrulewidth}
   	$\kappa_\tinySEC^\tinyWALL$ & Thermal conductivity of the secondary fluid pipe wall & 45 & W m\textsuperscript{-1} K\textsuperscript{-1} \\ \Xhline{2\arrayrulewidth}
   	$\mu_\tinyINT$ & Dynamic viscosity of the intermediate fluid & 0.066 -- 0.1174 & kg m\textsuperscript{-1} s\textsuperscript{-1} \\ 
   	\Xhline{2\arrayrulewidth}
   	$\mu_\tinySEC$ & Dynamic viscosity of the secondary fluid & 0.0073 -- 0.2213 & kg m\textsuperscript{-1} s\textsuperscript{-1} \\ 
   	\Xhline{2\arrayrulewidth}
   	$\rho_\tinyINT$ & Density of the intermediate fluid & 1080.1 -- 1113.2 & kg m\textsuperscript{-3} \\ 
   	\Xhline{2\arrayrulewidth}
   	$\rho_\tinyREF^\tinyLATMIN$ & Density of the two-phase refrigerant with zero vapour quality & 1291.60 & kg m\textsuperscript{-3} \\ 
   	\Xhline{2\arrayrulewidth}
   	$\rho_\tinyREF^\tinyLATMAX$ & Density of the two-phase refrigerant with one vapour quality & 6.76 & kg m\textsuperscript{-3} \\ 
   	\Xhline{2\arrayrulewidth}
   	$\rho_\tinyREFVAP$ & Density of the refrigerant in vapour phase & 6.40 -- 6.76 & kg m\textsuperscript{-3} \\ 
   	\Xhline{2\arrayrulewidth}
   	$\rho_\tinySEC$ & Density of the secondary fluid & 1040.7 -- 1068.4 & kg m\textsuperscript{-3} \\ 
   	\Xhline{2\arrayrulewidth}
   	$\sigma_\tinyREF$ & Surface tension of the refrigerant & 0.01336 & N m\textsuperscript{-1} \\ 
   	\Xhline{2\arrayrulewidth}
   	$\chi_\tinyREFTWO$ & Mean vapour quality of the refrigerant in two-phase zone& 0.7775 & p.u. \\ 
   	\Xhline{2\arrayrulewidth}
	\end{tabu}}
\end{table}

\begin{table} [htbp]
    \centering
    \caption{Convective heat transfer coefficients}
    \label{tabHeatTransferCoefficients}
    \scalebox{0.6}[0.6]{ \tabulinesep=0.5mm {}
        \begin{tabu} {L{1.5cm}  L{10cm} L{3.5cm} L{2cm}}
            \emph{Symbol} & \emph{Description} & \emph{Value} & \emph{Units} \\ 
            \Xhline{3\arrayrulewidth} 
            $\alpha_\tinyPCM^\tinyEXT$ & Convective heat transfer coefficient due to natural convection on the outer surface of the PCM capsule & 97.3 -- 144.7 & W m\textsuperscript{-2} K\textsuperscript{-1} \\ 
            \Xhline{2\arrayrulewidth}
            $\alpha_\tinyREFTWO^\tinyEXT$ & Convective heat transfer coefficient due to natural convection on the outer surface of the refrigerant pipes at the two-phase zone & 53.9 -- 142.3 & W m\textsuperscript{-2} K\textsuperscript{-1} \\ 
            \Xhline{2\arrayrulewidth}
            $\alpha_\tinyREFTWO^\tinyINT$ & Convective heat transfer coefficient due to forced convection on the inner surface of the refrigerant pipes when it is two-phase & 78.1 -- 429.4 & W m\textsuperscript{-2} K\textsuperscript{-1} \\ 
            \Xhline{2\arrayrulewidth}
            $\alpha_\tinyREFVAP^\tinyEXT$ & Convective heat transfer coefficient due to natural convection on the outer surface of the refrigerant pipes at the superheated zone & 46.3 -- 299.2 & W m\textsuperscript{-2} K\textsuperscript{-1} \\ 
            \Xhline{2\arrayrulewidth}
            $\alpha_\tinyREFVAP^\tinyINT$ & Convective heat transfer coefficient due to forced convection on the inner surface of the refrigerant pipes when it is superheated vapour & 1.2 -- 34.9 & W m\textsuperscript{-2} K\textsuperscript{-1} \\ 
            \Xhline{2\arrayrulewidth}
            $\alpha_\tinySEC^\tinyEXT$ & Convective heat transfer coefficient for natural convection in the exterior of the secondary fluid pipes & 67.6 -- 88.8 & W m\textsuperscript{-2} K\textsuperscript{-1} \\ \Xhline{2\arrayrulewidth}
            $\alpha_\tinySEC^\tinyINT$ & Convective heat transfer coefficient for internal forced convection of the secondary fluid & 40.4 -- 40.5 & W m\textsuperscript{-2} K\textsuperscript{-1} \\ 
            \Xhline{2\arrayrulewidth}
    \end{tabu}}
\end{table}


\section{Continuous modelling of the cold-thermal energy charge/discharge cycles}

\label{secSystemModelling}

Modelling and simulation are unavoidable steps in our work, in order to validate or refuse a given design proposal, prior to any actual modification of the existing refrigeration plant. In particular, this section addresses the modelling stage in due detail, describing the mathematical relationships and thermodynamic interactions between the different elements.

An important aspect of the continuous model to be presented in this section is that it comprises only a limited description of the system dynamics. Specifically, as we will see, it can only properly describe full charging or discharging operations. Nevertheless, it is important to set up such a model, for several reasons. By its own, it serves as a tool to verify that the specified full charging/discharging time periods are met, given a certain type of PCM or intermediate fluid and their respective volumes and properties. The continuous model to be presented is based on other works in the related literature, where the freezing and melting processes of a PCM capsule have been already studied and experimentally validated  \cite{bedecarrats2009bstudy,bedecarrats1996phase,ismail2009numerical,amin2014effective,temirel2017solidification}.

Besides, this simplified model is intended to be used as a validation instrument for the more complete and sophisticated discrete model. Through it we will be able to verify, for instance, that if we limit the operation of the discrete model to a full charging (or discharging) cycle, the behaviour of the discrete model draws near the prediction made by the continuous one, as the quantization level and sampling time of the discrete model is reduced.

Some assumptions have been made along the modelling process to be described, namely:

\begin{enumerate}[(a)]

\item The intermediate fluid is assumed to have, at each instant, a homogeneous temperature in the whole tank volume, due to its very high thermal conductivity. Hence, there is not spatial variation in the temperature of this fluid, but only time variation.

\item Supercooling is a well-known phenomenon in the context of phase change materials \cite{streicher2008simulation}. The basic principle is that solidification begins only after the temperature of the fluid is lowered beyond the freezing point. This phenomenon is expected to occur during the initial stages of solidification. Since our study is essentially focused on PCM working in latent zone, without specifically dealing the early stages of solidification, supercooling effect is neglected.

\item Another feature of homogeneous mixtures is that phase change occurs in a narrow temperature range, instead in a specific temperature value, unlike what is expected in pure substances. In the case of PCM, in particular, this circumstance is aggravated as the temperature range is different, depending on whether the material is experiencing a freezing process or a melting process. This causes the presence of a hysteresis phenomenon between charge and discharge. This effect has been overlooked in our study \cite{streicher2008simulation}.

\end{enumerate}

In the continuous model, the behaviour of the PCM capsules is simplified as follows.

\subsection{General description of the charging state}

During the charging cycle, it is assumed that only one single inward freezing front is present, as expected during a continuous charging cycle starting from a discharged steady state. The previous assumption of full charge/discharge cycles, justifies this simplification. The discrete model to be described later will be able to overcome this limitation.

\begin{figure}[htbp]
    \centerline{\includegraphics[width=10cm,trim = 0 33 0 33,clip]
    {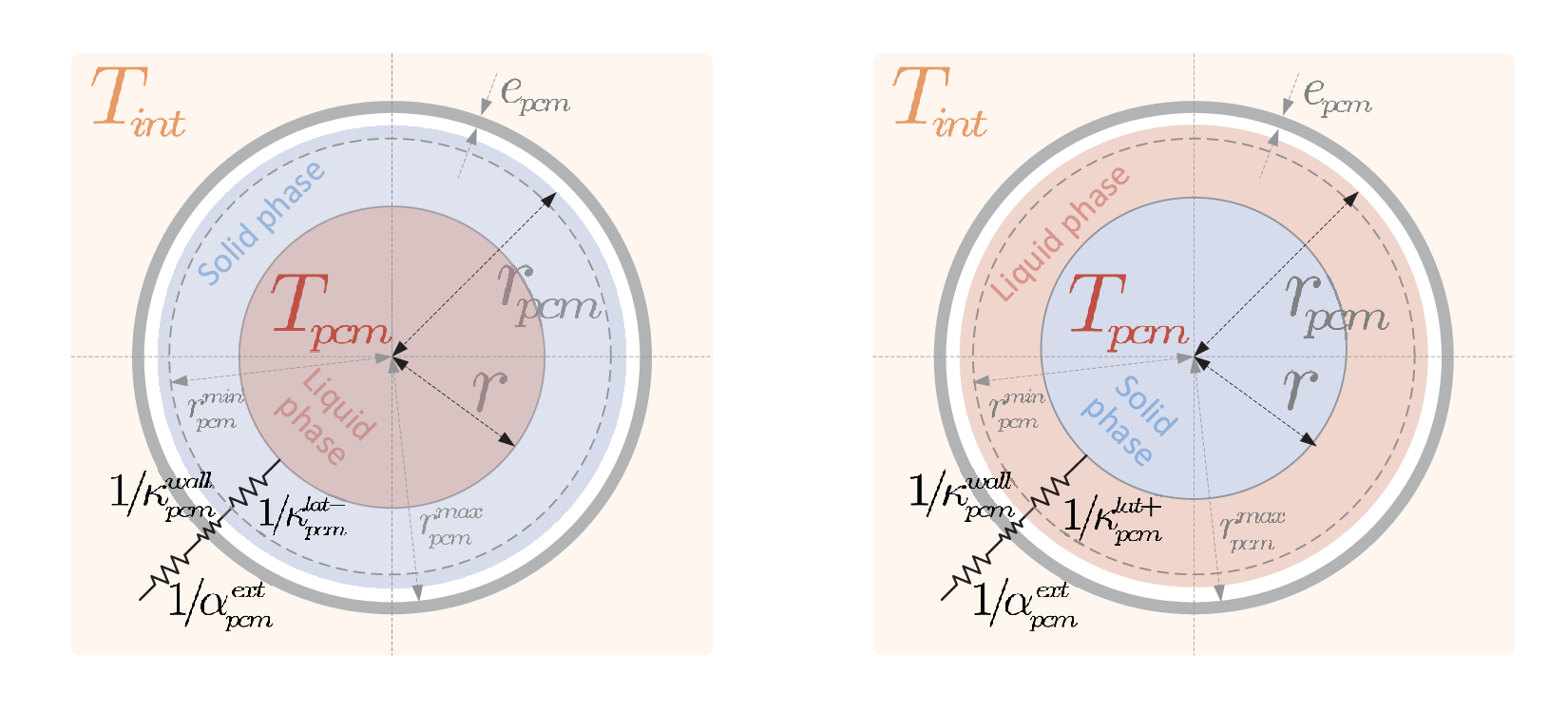}}
    \caption{Schematic picture of the PCM capsule during charging/discharging cycle (left/right, respectively). Electric analogy of the thermal resistances is shown.}
    \label{figPCMcapsule}
\end{figure}

The left part of Figure \ref{figPCMcapsule} shows the continuous model of the PCM capsule during the charging cycle. It can be seen that $r(t)$ is defined as the instantaneous radius of the spherical boundary between the solid and liquid PCM phases inside the capsule. Moreover, $r_{\!\tinyPCM}(t)$ corresponds to the instantaneous radius of the PCM capsule, taking into account that, due to the density difference between the frozen and melted PCM, the mass confined in the capsule is expected to reduce its volume. $r_{\!\tinyPCM}^{\,\tinyMAX}$ refers to the radius of the PCM capsule when it is completely melted (for instance, at the initial discharged steady state), and $r_{\!\tinyPCM}^{\,\tinyMIN}$ corresponds to the radius of the PCM capsule when it is completely frozen, achievable at the end of the charging cycle. 

As illustrated in the figure, a receding melted core and an inward-growing frozen external spherical shell are present. Thereby, the internal thermal energy stored inside each PCM capsule can be decomposed in two parts: the energy stored in the capsule core and the energy stored in the external spherical shell, as indicated in Equation \eqref{eq:EnergyPCM_ChargeAbsolute}.

\begin{equation}
   U_\tinyPCM^\tinyCHARGE(t) \,=\, \left[\rho_\tinyPCM^\tinyLATMAX \,
   h_\tinyPCM^\tinyLATMAX \frac{4\pi}{3}\, r(t)^3\right] \,+\, \left[\rho_\tinyPCM^\tinyLATMIN \,
   h_\tinyPCM^\tinyLATMIN \frac{4\pi}{3}\, (r_{\!\tinyPCM}(t)^3-r(t)^3)\right]
   \label{eq:EnergyPCM_ChargeAbsolute}
\end{equation}

Thermal resistances depicted in the figure are shown as dependent on their corresponding thermal conductivities or heat transfer coefficients, and they rule heat transfer by temperature difference between the PCM core and the intermediate fluid ($T_\tinyPCM(t) - T_\tinyINT(t)$), as will be detailed later in this section.

According to our assumption for the continuous model, the system has been fully discharged before each charging cycle, $r(t\!=\!0)=r_{\!\tinyPCM}(t\!=\!0)=r_{\!\tinyPCM}^{\,\tinyMAX}$, and, as soon as $r(t)$ reaches 0, the charging cycle is completed. The meaning of the electrical resistances in Figure \ref{figPCMcapsule}, used for the heat transfer computation between the intermediate fluid and the sphere core, will be discussed in Subsection \ref{sec:PCM_intermediate}.

\subsection{General description of the discharging state}

Only one single inward melting front is expected during discharging cycles, assuming again a continuous discharging cycle starting from a fully charged state. The right part of Figure \ref{figPCMcapsule} shows the behaviour of the PCM capsule during the discharging cycle, according to the continuous model. In this case, the core is frozen and the external shell is melted, again, presumably on the limits of the latent phase. The energy stored inside the PCM capsule is computed as shown in Equation \eqref{eq:EnergyPCM_DischargeAbsolute}.

\begin{equation}
   U_\tinyPCM^\tinyDCHARGE(t) \,=\, \left[\rho_\tinyPCM^\tinyLATMIN \, h_\tinyPCM^\tinyLATMIN \frac{4\pi}{3}\, r(t)^3\right] \,+\, \left[\rho_\tinyPCM^\tinyLATMAX \, h_\tinyPCM^\tinyLATMAX \frac{4\pi}{3}\, (r_{\!\tinyPCM}(t)^3-r(t)^3)\right]
   \label{eq:EnergyPCM_DischargeAbsolute}
\end{equation}

In this case, $r(t\!=\!0)=r_{\!\tinyPCM}(t\!=\!0)=r_{\!\tinyPCM}^{\,\tinyMIN}$, and when the radius of the melting front becomes zero, $r(t)=0$, the discharging cycle is completed.

\subsection{\emph{Charge ratio} estimation in the continuous model}

During typical charging and discharging operations, it is assumed that the PCM is working on the latency limits. Our convention is defining the minimum thermal energy inside each PCM capsule as the energy stored just when the whole capsule reaches minimum enthalpy and thus minimum volume. Similarly, the maximum thermal energy inside each PCM capsule is defined as the energy stored just when the whole capsule reaches maximum enthalpy and thus maximum volume, as indicated in Equation Set \eqref{eq:maxminEnergyPCM}.

\begin{equation}
	\begin{aligned}
	U_\tinyPCM^\tinyMIN \,&=\, \rho_\tinyPCM^\tinyLATMIN \, h_\tinyPCM^\tinyLATMIN \frac{4\pi}{3}\, {r_{\!\tinyPCM}^{\,\tinyMIN}}^3 \\
	U_\tinyPCM^\tinyMAX \,&=\, \rho_\tinyPCM^\tinyLATMAX \, h_\tinyPCM^\tinyLATMAX \frac{4\pi}{3}\, {r_{\!\tinyPCM}^{\,\tinyMAX}}^3 \\
	\end{aligned}
	\label{eq:maxminEnergyPCM}
\end{equation}

By replacing Equations \eqref{eq:EnergyPCM_ChargeAbsolute} and \eqref{eq:EnergyPCM_DischargeAbsolute} in Equation \eqref{eq:ChargeRatioDefinition2}, the \emph{charge ratio} during both cycles is computed as indicated in Equation Set \eqref{eq:ChargeRatioDefinition_ext}.

\begin{equation}
	\begin{aligned}
	\gamma_\tinyTES^\tinyCHARGE(t) \,&=\, 
	\dfrac{\rho_\tinyPCM^\tinyLATMAX \, h_\tinyPCM^\tinyLATMAX \, {r_{\!\tinyPCM}^{\,\tinyMAX}}^3 \,-\, \rho_\tinyPCM^\tinyLATMIN \, h_\tinyPCM^\tinyLATMIN \, {r_\tinyPCM (t)}^3 \,-\, (\rho_\tinyPCM^\tinyLATMAX \, h_\tinyPCM^\tinyLATMAX \,-\, \rho_\tinyPCM^\tinyLATMIN \, h_\tinyPCM^\tinyLATMIN) \, {r(t)}^3} {\rho_\tinyPCM^\tinyLATMAX \, h_\tinyPCM^\tinyLATMAX \, {r_{\!\tinyPCM}^{\,\tinyMAX}}^3 \,-\, \rho_\tinyPCM^\tinyLATMIN \, h_\tinyPCM^\tinyLATMIN \, {r_{\!\tinyPCM}^{\,\tinyMIN}}^3} 
	\\
	\gamma_\tinyTES^\tinyDCHARGE(t)\,&=\,
	\dfrac{\rho_\tinyPCM^\tinyLATMAX \, h_\tinyPCM^\tinyLATMAX \, {r_{\!\tinyPCM}^{\,\tinyMAX}}^3 \,-\, \rho_\tinyPCM^\tinyLATMIN \, h_\tinyPCM^\tinyLATMIN \, {r_\tinyPCM (t)}^3 \,+\, (\rho_\tinyPCM^\tinyLATMAX \, h_\tinyPCM^\tinyLATMAX \,-\, \rho_\tinyPCM^\tinyLATMIN \, h_\tinyPCM^\tinyLATMIN) \, {r(t)}^3} {\rho_\tinyPCM^\tinyLATMAX \, h_\tinyPCM^\tinyLATMAX \, {r_{\!\tinyPCM}^{\,\tinyMAX}}^3 \,-\, \rho_\tinyPCM^\tinyLATMIN \, h_\tinyPCM^\tinyLATMIN \, {r_{\!\tinyPCM}^{\,\tinyMIN}}^3} 
	\\
	\end{aligned}
	\label{eq:ChargeRatioDefinition_ext}
\end{equation}

Since $r(t)$ is defined as the inward radius of the freezing front during charging operation or the inward radius of the melting front during discharging operation, our definition of $\gamma_\tinyTES(t)$ is particular for each mode. Of course, this definition is only valid during operation of the PCM in latent state. In sensible zone, $r(t)$ is not relevant, since the whole PCM capsule is either subcooled solid or superheated liquid. According to this, the system with subcooled frozen PCM  or superheated liquid PCM should be represented by an output value $\gamma_\tinyTES(t)>1$ or $\gamma_\tinyTES(t)<0$, respectively.

\subsection{Intermediate fluid dynamics}

The intermediate fluid is completely still inside the TES tank, at constant atmospheric pressure, though closed, bathing the PCM capsules and the refrigerant and secondary fluid bundles of tubes. Its composition is a glycol-water mixture with very high thermal conductivity, providing an efficient thermal transfer between the refrigerant and the PCM capsules, during the charging cycle, or between the PCM capsules and the secondary fluid, during the discharging cycle. Consequently,
the intermediate fluid is assumed to have, at each instant, a homogeneous temperature in the whole tank volume.

The energy-balance equation of the intermediate fluid is shown in Equation \eqref{eqPrincipal}, where the main items are detailed below.

\begin{equation}
	-m_\tinyINT\,c_{\!p_\tinyINT} \frac{dT_\tinyINT(t)}{dt} \,=\, n_\tinyREF \, \dot
	Q_\tinyREF(t) + n_\tinySEC \, \dot Q_\tinySEC(t) + n_\tinyPCM \, \dot Q_\tinyPCM(t) + \dot
	Q_\tinyENV(t) 
	\label{eqPrincipal}
\end{equation}

\begin{itemize}
   \item $T_\tinyINT(t)$: Temperature of the intermediate fluid (assumed to be instantaneously homogeneous inside the intermediate fluid reservoir).
   \item $\dot Q_\tinyREF(t)$: Cooling power transferred from each single refrigerant pipe to the intermediate fluid (typically positive).
   \item $\dot Q_\tinySEC(t)$: Cooling power transferred from each single secondary fluid pipe to the intermediate fluid (typically negative).
   \item $\dot Q_\tinyPCM(t)$: Cooling power transferred from each PCM capsule to the intermediate fluid (positive during discharging cycle, negative during charging cycle).
   \item $\dot Q_\tinyENV(t)$: Cooling power transferred from the environment to the intermediate (typically negative). For the sake of simplicity, it will be assumed that the tank is perfectly insulated and does not exchange heat with the environment: $\dot Q_\tinyENV(t)=0$.
\end{itemize}

On the one hand, during the charging cycle, all secondary mass flow is stopped ($\dot Q_\tinySEC(t)=0$), the refrigerant is providing cooling power to the intermediate fluid ($\dot Q_\tinyREF(t)>0$) while each one of the PCM capsules is contributing to store this cold-energy, taking it from the intermediate fluid ($\dot Q_\tinyPCM<0$).

On the other hand, during the discharging cycle, the refrigerant mass flow is stopped ($\dot Q_\tinyREF(t)=0$), the PCM capsules transfer cold-energy to the intermediate fluid ($\dot Q_\tinyPCM(t)>0$), while the secondary fluid is taking cooling power from the intermediate fluid ($\dot Q_\tinySEC(t)<0$).


\subsection{PCM-intermediate fluid dynamics} \label{sec:PCM_intermediate}

It is assumed that the cooling power transferred between the intermediate fluid and each PCM capsule is completely devoted to phase change of the spherical layer closest to the boundary between the solid and liquid phases inside the capsule. Then, during the charging cycle, $\dot Q_\tinyPCM$ is devoted to freezing the very first layer of the liquid core of the PCM capsule, as indicated in Equation \eqref{eq:CoolingPowerCharge}. Furthermore, during the discharging cycle, $\dot Q_\tinyPCM$ is devoted to melting the very first layer of the frozen core of the PCM capsule, as shown in Equation \eqref{eq:CoolingPowerDischarge}.

\begin{equation}
	\dot Q_\tinyPCM(t) \,=\, \rho_\tinyPCM ^\tinyLATMAX \, h_\tinyPCM^\tinyLAT 4\pi\, r(t)^2 \,\dot r(t)\; \\
	\label{eq:CoolingPowerCharge} 
\end{equation}

\begin{equation}
	\dot Q_\tinyPCM(t) \,=\, \,-\, \rho_\tinyPCM ^\tinyLATMIN\, h_\tinyPCM^\tinyLAT 4\pi\, r(t)^2 \,\dot r(t)\; \\
	\label{eq:CoolingPowerDischarge}
\end{equation}

It should be clear at this point that $\dot r(t)<0$ during both cycles. Furthermore, applying mass conservation inside the PCM capsule, it is possible to derive the relationship between $\dot r(t)$ and $\dot r_\tinyPCM(t)$ shown in Equation \eqref{eq:r_PCM_Charge} regarding the charging cycle, and in Equation \eqref{eq:r_PCM_Discharge} concerning the discharging cycle.

\begin{equation}
	\dot r_\tinyPCM(t) \,=\, \dfrac{\rho_\tinyPCM ^\tinyLATMIN \,-\, \rho_\tinyPCM ^\tinyLATMAX}{\rho_\tinyPCM ^\tinyLATMIN} \, \dfrac{{r(t)}^2}{{r_\tinyPCM(t)}^2} \, \dot r(t) \, \,<\, 0 \\
	\label{eq:r_PCM_Charge} 
\end{equation}

\begin{equation}
	\dot r_\tinyPCM(t) \,=\, \,-\, \dfrac{\rho_\tinyPCM ^\tinyLATMIN \,-\, \rho_\tinyPCM ^\tinyLATMAX}{\rho_\tinyPCM ^\tinyLATMAX} \, \dfrac{{r(t)}^2}{{r_\tinyPCM(t)}^2} \, \dot r(t) \, \,>\, 0\\
	\label{eq:r_PCM_Discharge} 
\end{equation}

Heat transfer between the intermediate fluid and the PCM capsules is ruled by Equation \eqref{eq:Q_PCM_main}, or in other terms (convenient for heat transfer computation), by Equation \eqref{eq:Q_PCM_aux}.

\begin{equation}
	\dot Q_\tinyPCM(t) \,=\, \frac{T_\tinyINT(t) - T_\tinyPCM(t)}{R_\tinyPCM(t)} \,=\, \frac{T_\tinyINT(t) - T_\tinyPCM^\tinyLAT}{R_\tinyPCM^\tinyCONDINT(t) + R_\tinyPCM^\tinyCONDWALL + R_\tinyPCM^\tinyCONVEXT(t)}
	\label{eq:Q_PCM_main} 
\end{equation}

\begin{equation}
	\dot Q_\tinyPCM(t) \,=\, \frac{T_\tinyPCM^\tinyWALL(t) - T_\tinyPCM^\tinyLAT}{R_\tinyPCM^\tinyCONDINT(t) + R_\tinyPCM^\tinyCONDWALL} 
	\label{eq:Q_PCM_aux} 
\end{equation}

Some details regarding the terms appearing in Equations \eqref{eq:Q_PCM_main} and \eqref{eq:Q_PCM_aux} are provided below:

\begin{itemize}
	
	\item $T_\tinyPCM(t)$: Temperature of the PCM core, which matches $T_\tinyPCM^\tinyLAT$ when working in latent zone. The external spherical shell acts as a mere thermal conduction layer between the capsule core and the intermediate fluid.
	
	\item $R_\tinyPCM(t)$: Global thermal resistance defining heat transfer between the PCM core and the intermediate fluid, composed by the following three terms arranged in series:
	
	\begin{itemize}
		
		\item[$\circ$] $R_\tinyPCM^\tinyCONDINT(t)$: Thermal resistance related to heat conduction inside the spherical PCM shell, located between the capsule core and the polymer inner surface. It is computed as shown in Equation \eqref{eq:R_PCM_CONDINT}.
		
		\begin{equation}
			R_\tinyPCM^\tinyCONDINT(t) \,=\, \frac{r_{\!\tinyPCM}(t) - r(t)} {\kappa_\tinyPCM\,4\pi\,r_{\!\tinyPCM}(t)\,r(t)} \,=\, \frac{1}{\kappa_\tinyPCM\,4\pi}\left(\frac{1}{r(t)} - \frac{1}{r_{\!\tinyPCM}(t)}\right) 
			\label{eq:R_PCM_CONDINT} 
		\end{equation}
		
		As shown in the Figure \ref{figPCMcapsule}, on the one hand, during charging operation $\kappa_\tinyPCM=\kappa_\tinyPCM^\tinyLATMIN$, since the external spherical shell is frozen. On the other hand, during discharging operations it has been studied in the literature that, due to buoyancy forces caused by density difference between the solid and liquid, natural convection appears within the capsule when the PCM melts. It boosts heat transfer and accelerates the PCM melting. This phenomenon has been modelled in the literature by considering a pure thermal conduction model with an \emph{effective} thermal conductivity, $\kappa_{\tinyPCM,\,\tinyEFF}^{\tinyLATMAX}$, higher than that of the PCM liquid \cite{amin2014effective}. Indeed, experimental correlations according to the \emph{Rayleigh} number state that $\kappa_{\tinyPCM,\,\tinyEFF}^{\tinyLATMAX}$ might be between two and three times higher than $\kappa_\tinyPCM^\tinyLATMAX$. This phenomenon has been also considered in the presented modelling.      
		
		\item[$\circ$] $R_\tinyPCM^\tinyCONDWALL$: Thermal resistance related to heat conduction of the PCM capsule polymer coating, computed as indicated in Equation \eqref{eq:R_PCM_CONDWALL} according to the spherical geometry.
		
		\begin{equation}
			R_\tinyPCM^\tinyCONDWALL \,=\, \frac{1}{\kappa_\tinyPCM^\tinyWALL 4\pi}\left(\frac{1}{r_{\!\tinyPCM}^{\,\tinyMAX}} - \frac{1}{r_{\!\tinyPCM}^{\,\tinyMAX} \!\!+\!e_\tinyPCM}\right) 
			\label{eq:R_PCM_CONDWALL} 
		\end{equation}
		
		\item[$\circ$] $R_\tinyPCM^\tinyCONVEXT$: Thermal resistance related to natural convection of the intermediate fluid around the PCM capsule, computed as shown in Equation \eqref{eq:R_PCM_CONVEXT}. The heat transfer coefficient, $\alpha_\tinyPCM^\tinyEXT(t)$, is computed from the \emph{Nusselt} number, using the correlation for external natural convection in spheres (detailed in \ref{secAppendixExternalConvectionSpheres}).
		
		\begin{equation}
			R_\tinyPCM^\tinyCONVEXT(t) \,=\, \frac{1}{\alpha_\tinyPCM^\tinyEXT(t)\,4\pi\, (r_{\!\tinyPCM}^{\,\tinyMAX}\!\!+\!e_\tinyPCM)^2} 
			\label{eq:R_PCM_CONVEXT} 
		\end{equation}
		
	\end{itemize}
	
\end{itemize}


\subsection{Refrigerant - intermediate fluid dynamics}

\subsubsection{Notation for the \emph{effectiveness-NTU} method}

The \emph{effectiveness-NTU} method is widely used to describe the heat transfer in some configurations of heat exchangers \cite{holman2001heat,bergman2011fundamentals,ESDU}.

We denote, in general, $T_\tinyHOT^\tinyIN(t)$ and $T_\tinyCOLD^\tinyIN(t)$ the inlet temperature of the hot and cold fluid streams in the heat exchanger, respectively. The respective heat capacity rates are $C_\tinyCOLD(t) \,=\, \dot m_\tinyCOLD(t)\,c_{\!p_\tinyCOLD}$ and $C_\tinyHOT(t) \,=\,
\dot m_\tinyHOT(t)\,c_{\!p_\tinyHOT}$. From these, $C\,^\tinyMIN(t)$ and $C\,^\tinyMAX(t)$ are defined as shown in Equation Set \eqref{eq:eNTU_Cmax_Cmin}.

\begin{equation}
	\begin{aligned}
	C\,^\tinyMIN(t) &= \min\left(C_\tinyCOLD(t),C_\tinyHOT(t)\right)\, \\
    C\,^\tinyMAX(t) &= \max\left(C_\tinyCOLD(t),C_\tinyHOT(t)\right) \\
    \end{aligned}
    \label{eq:eNTU_Cmax_Cmin} 
\end{equation}

The steady-state model of the cold thermal power transferred between both fluids, can be described by Equation \eqref{eq:eNTUgeneral}.

\begin{equation}
   \dot Q(t) \,=\, \varepsilon(t)\, C\,^\tinyMIN(t)\,(T_\tinyHOT^\tinyIN(t) - T_\tinyCOLD^\tinyIN(t))
   \label{eq:eNTUgeneral}
\end{equation}

The effectiveness, $\varepsilon(t)$, is defined as shown in Equation Set \eqref{eq:eNTU_epsilon}, where $R(t)$ represents the global thermal resistance in the heat exchanger.

\begin{equation}
	\begin{aligned}
    \varepsilon(t) \,&=\, f(\NTU,\,C\,^{\!\tinyRAT}) \\
    \NTU(t) \,&=\, \dfrac{1}{R(t)\,C\,^\tinyMIN(t)} \\
    C\,^{\!\tinyRAT}(t) \,&=\, \dfrac{C\,^\tinyMIN(t)}{C\,^\tinyMAX(t)} \\
    \end{aligned}
	\label{eq:eNTU_epsilon} 
\end{equation}

The expression of the function $f(N\!T\!U,\,C\,^{\!\tinyRAT})$, in the case of parallel-flow heat exchangers, is described in Equation \eqref{eqEffectivenessParallelCurrent}, whereas in the case of counter-current flow heat exchangers, this expression becomes that included in Equation \eqref{eqEffectivenessCounterCurrent}.

\begin{equation}
   f(\NTU,\,C\,^{\!\tinyRAT}) \,=\,\dfrac{1-e^{-\NTU\,(1-C\,^{\!\tinyRAT})}}{1+C\,^{\!\tinyRAT}}
   \label{eqEffectivenessParallelCurrent}
\end{equation}

\begin{equation}
   f(\NTU,\,C\,^{\!\tinyRAT}) \,=\,\dfrac{1-e^{-\NTU\,(1-C\,^\tinyRAT)}}{1-C\,^{\!\tinyRAT}\,e^{-\NTU(1-C\,^\tinyRAT)}}
   \label{eqEffectivenessCounterCurrent}
\end{equation}

\subsubsection{Heat transfer between the refrigerant and the intermediate fluid}

During the charging cycle, the refrigerant flows through a bundle of pipes, bathed in the intermediate fluid, inside the TES tank. The refrigerant is fed at a certain pressure, according to the operating conditions of the refrigeration plant.

The \emph{moving boundary} approach \cite{Rasmussen2005,Liang2010} is applied to
model the refrigerant behaviour along its way through the tank. By design, the refrigerant enters the tank as two-phase fluid and it usually comes out as slightly superheated vapour (\emph{mode 1}), but, alternatively, it may also come out as two-phase fluid (\emph{mode 2}).

\begin{figure*}[htbp]
    \centerline{\includegraphics[width=15cm,trim = 0 20 0 20,clip]
    {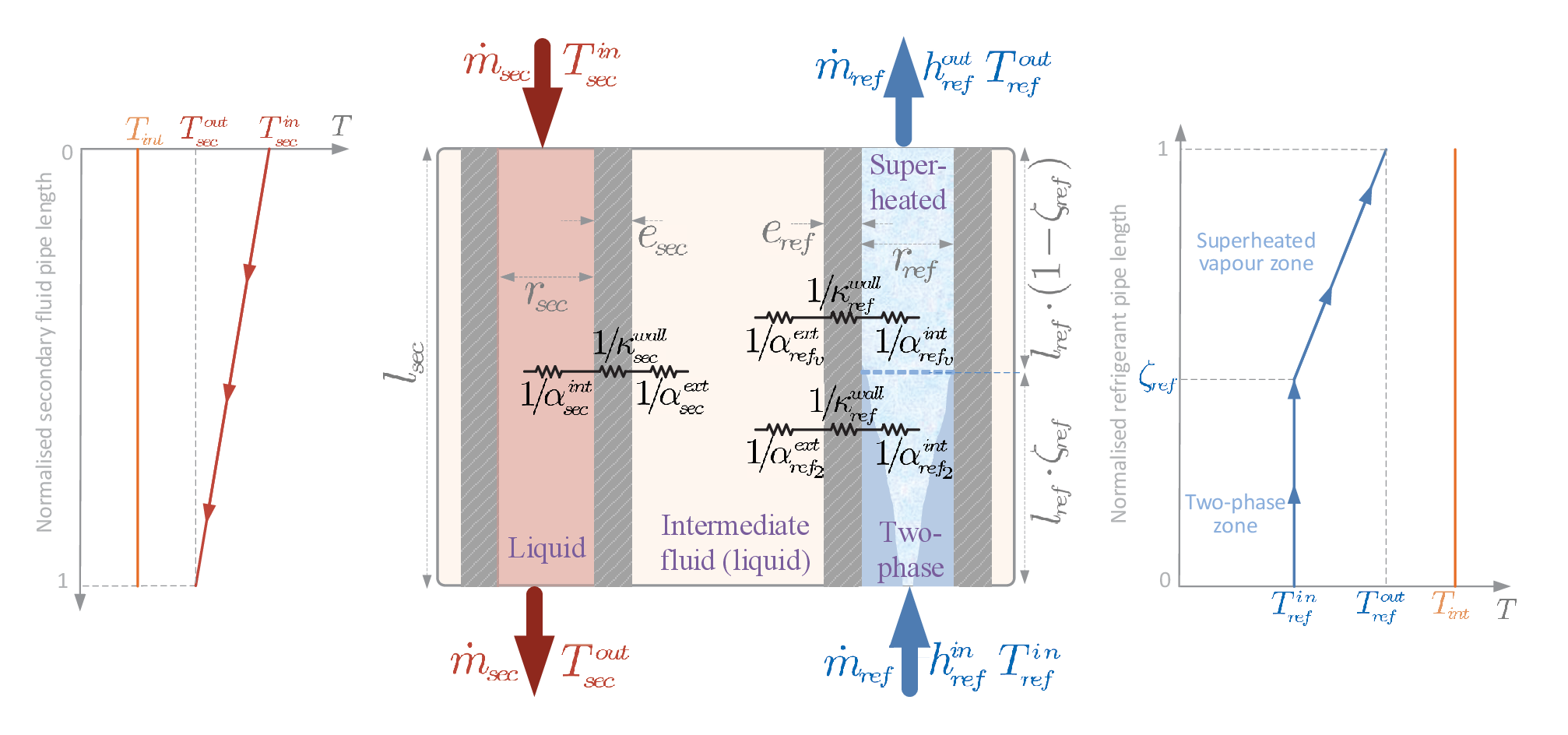}}
    \caption{Schematic picture of the thermal exchange along refrigerant and secondary fluid single pipes.}
    \label{figHeatExchangeTubes}
\end{figure*}

A diagram of the heat transfer inside the intermediate fluid tank is shown in Figure
\ref{figHeatExchangeTubes}. Regarding the refrigerant, \emph{mode 1} is depicted, being the pipe divided into two variable-length zones, where the refrigerant is two-phase fluid or superheated vapour, respectively.

According to this configuration, expressions describing heat transfer between the refrigerant and the intermediate fluid need to be provided for both zones, the superheated vapour zone and the two-phase zone, since the global cooling power transferred is computed as shown in Equation \eqref{eq:QREF}.

\begin{equation}
	\dot Q_\tinyREF(t) \,=\, \dot Q_\tinyREFTWO(t) + \dot Q_\tinyREFVAP(t) 
	\label{eq:QREF}
\end{equation}

The right part of Figure \ref{figHeatExchangeTubes} represents the qualitative temperature diagram of the refrigerant and intermediate fluid, as they exchange heat along the refrigerant pipes. By comparison with the refrigerant and the secondary fluid, the intermediate fluid can be considered a very large mass, that instantaneously attains homogeneous temperature in the whole tank volume. This is the reason why the intermediate fluid temperature exhibits no spatial gradient in the figure.

In the case of the two-phase zone, the involved fluids, i.e. the intermediate fluid and the refrigerant in latent phase, is each one at instantaneously homogeneous temperature along the thermal transfer area; hence, mere integration of the general heat transfer expression suffices, as described in Equation \eqref{eq:QREFTWO}.

\begin{equation}
	\dot Q_\tinyREFTWO(t) \,=\, \frac{T_\tinyINT(t) - T_\tinyREF^\tinyIN(t)}{R_\tinyREFTWO(t)} \,=\, \frac{T_\tinyINT(t) - T_\tinyREF^\tinyIN(t)}{R_\tinyREFTWO^\tinyCONVINT\!(t)  + R_\tinyREFTWO^\tinyCONDWALL\!(t)  + R_\tinyREFTWO^\tinyCONVEXT\!(t)} 
	\label{eq:QREFTWO}
\end{equation}

$R_\tinyREFTWO(t)$ represents the global thermal resistance in the heat transfer process between the refrigerant and intermediate fluid, regarding the two-phase zone, where three thermal resistances are arranged in series. The conduction term of the global resistance is given by the cylindrical geometry and thermal conductivity of the refrigerant pipe wall, while the internal and external convective terms (forced and natural, respectively) of the global resistance are computed considering the length of the refrigerant two-phase zone and the corresponding convective heat transfer coefficient, as shown in Equation Set \eqref{eq:RREFTWO}.

\begin{equation}
	\begin{aligned}
	R_\tinyREFTWO^\tinyCONDWALL\!(t) \,&=\, \frac{\ln{\frac{r_{\!\tinyREF}+e_\tinyREF}{r_{\!\tinyREF}}}} {\kappa_\tinyREF^\tinyWALL\,2\pi\,l_\tinyREF\,\zeta_\tinyREF\!(t)} \\
    R_\tinyREFTWO^\tinyCONVINT\!(t) \!&=\! \frac{1}{\alpha_\tinyREFTWO^\tinyINT(t)\,2\pi\,r_{\!\tinyREF}\,l_\tinyREF\,\zeta_\tinyREF\!(t)}\; \\
    R_\tinyREFTWO^\tinyCONVEXT(t) \!&=\! \frac{1}{\alpha_\tinyREFTWO^\tinyEXT(t)\,2\pi\,(r_{\!\tinyREF}\!\!+\!e_\tinyREF)\,l_\tinyREF\,\zeta_\tinyREF\!(t)}
   \end{aligned}
   \label{eq:RREFTWO}
\end{equation}

On the one hand, Klimenko's method is used to compute the internal convective coefficient in the two-phase zone, $\alpha_\tinyREFTWO^\tinyINT(t)$, which is described in \ref{secAppendixInternalForcedConvectionPipesKlimenko}. On the other hand, $\alpha_\tinyREFTWO^\tinyEXT(t)$ is obtained through the correlation for external natural convection in vertical tubular pipes, described also in \ref{secAppendixExternalConvectionPipes}.

An additional equation is needed in this case, in order to solve for
$\zeta_\tinyREF(t)$, which is described in Equation \eqref{eq:QREFTWO_add}.

\begin{equation}
	\dot Q_\tinyREFTWO(t) \,=\, \dot m_\tinyREF\!(t)\, (h_\tinyREF^\tinyLATMAX - h_\tinyREF^\tinyIN(t))
	\label{eq:QREFTWO_add}
\end{equation}

Furthermore, the \emph{effectiveness-NTU} method is used to describe heat transfer in the superheated vapour zone, where the intermediate fluid acts as the hot source, while the refrigerant is the cold source. As mentioned before, the intermediate fluid is completely still and considered, by comparison, a very large mass with instantaneously homogeneous temperature. According to this, its mass flow rate is approximated by $C_\tinyINT\approx\infty$; thus, $C\,^\tinyMIN \approx C_\tinyREFVAP$, being $C_\tinyREFVAP(t)\,=\, c_{\!p_{\,\tinyREFVAP}}\,\dot
m_\tinyREF(t)$, and $C\,^\tinyRAT\approx 0$.

Equation Set \eqref{eq:eNTUgeneral}-\eqref{eq:eNTU_epsilon} then becomes Equation Set \eqref{eq:eNTU_epsilon_REFVAP}, while both expressions \eqref{eqEffectivenessParallelCurrent} and \eqref{eqEffectivenessCounterCurrent} collapse into that shown in Equation \eqref{eq:fNTU_REFVAP}.

\begin{equation}
	\begin{aligned}
   \dot Q_\tinyREFVAP(t) \,&=\, \varepsilon_\tinyREFVAP(t)\, C_\tinyREFVAP(t)\,(T_\tinyINT(t) - T_\tinyREF^\tinyIN(t)) \\
   \varepsilon_\tinyREFVAP(t) \,&=\, f(\NTU_\tinyREFVAP) \\
   \NTU_\tinyREFVAP(t) \,&=\,  \dfrac{1}{R_\tinyREFVAP(t)\,C_\tinyREFVAP(t)}   
   \end{aligned}
   \label{eq:eNTU_epsilon_REFVAP}
 \end{equation}
 
\begin{equation}
	f(\NTU) \,=\,1-e^{-\NTU} 
	\label{eq:fNTU_REFVAP}
\end{equation}

$R_\tinyREFVAP(t)$ represents the global thermal resistance in the heat transfer process between the refrigerant and the intermediate fluid, in the superheated vapour zone. The three terms indicated in Equation \eqref{eq:R_REFVAP} are involved, whereas their expressions are shown in Equation Set \eqref{eq:R_REFVAP_DESCRIPTION}.

\begin{equation}
	R_\tinyREFVAP\!(t) \,=\, R_\tinyREFVAP^\tinyCONVINT\!(t)  + R_\tinyREFVAP^\tinyCONDWALL\!(t)  + R_\tinyREFVAP^\tinyCONVEXT\!(t) 
	\label{eq:R_REFVAP}
\end{equation}

\begin{equation}
	\begin{aligned}
	R_\tinyREFVAP^\tinyCONDWALL\,(t) \,&=\, \frac{\ln{\frac{r_{\!\tinyREF}+e_\tinyREF}{r_{\!\tinyREF}}}} {\kappa_\tinyREF^\tinyWALL\,2\pi\,l_\tinyREF\,(1\!-\!\zeta_\tinyREF\!(t))} \\
	R_\tinyREFVAP^\tinyCONVINT\,(t) \!&=\! \frac{1}{\alpha_\tinyREFVAP^\tinyINT(t)\,2\pi\,r_{\!\tinyREF}\,l_\tinyREF\,(1\!-\!\zeta_\tinyREF\!(t))}\; \\
	R_\tinyREFVAP^\tinyCONVEXT(t) \!&=\! \frac{1}{\alpha_\tinyREFVAP^\tinyEXT(t)\,2\pi\,(r_{\!\tinyREF}\!\!+\!e_\tinyREF)\,l_\tinyREF\,(1\!-\!\zeta_\tinyREF\!(t))} 
	\end{aligned}
	\label{eq:R_REFVAP_DESCRIPTION}
\end{equation}

The correlation for external natural convection in vertical tubular pipes is used again to compute the external convective coefficient in the superheated vapour zone, $\alpha_\tinyREFVAP^\tinyEXT(t)$, as described in \ref{secAppendixExternalConvectionPipes}. Moreover, for the internal convective coefficient, $\alpha_\tinyREFVAP^\tinyINT(t)$, the correlation for internal forced convection in vertical tubular pipes (described in \ref{secAppendixInternalForcedConvectionPipes}) is used.

\subsection{Secondary fluid - intermediate fluid dynamics}

During the discharging cycle, the secondary fluid is always in liquid phase, flowing through a bundle of tubes, at constant pressure, submerged in the still intermediate fluid. The left part of Figure \ref{figHeatExchangeTubes} represents the qualitative temperature diagram of heat transfer between the secondary and intermediate fluids. The \emph{effectiveness-NTU} method is used again, where the secondary fluid acts as the hot source, while the intermediate fluid is the cold source. Its mass flow rate is approximated by $C_\tinyINT\approx\infty$; thus, $C\,^\tinyMIN \approx C_\tinySEC$, $C_\tinySEC(t)=c_{p_{\,\tinySEC}}\,\dot m_\tinySEC(t)$, and $C\,^\tinyRAT\approx 0$.

Equation Set \eqref{eq:eNTUgeneral}-\eqref{eq:eNTU_epsilon} then becomes Equation Set \eqref{eq:eNTU_epsilon_SEC}, while once again both expressions \eqref{eqEffectivenessParallelCurrent} and \eqref{eqEffectivenessCounterCurrent} collapse into that indicated in Equation \eqref{eq:fNTU_REFVAP}.

\begin{equation}
	\begin{aligned}
   	\dot Q_\tinySEC(t) \,&=\, \varepsilon_\tinySEC(t)\, C_\tinySEC(t)\,(T_\tinyINT(t) -  T_\tinySEC^\tinyIN(t)) \\
   	\varepsilon_\tinySEC(t) \,&=\, f(\NTU_\tinySEC)  \\
   	\NTU_\tinySEC(t) \,&=\,  \dfrac{1}{R_\tinySEC(t)\,C_\tinySEC(t)} \\
 	\end{aligned}
 	\label{eq:eNTU_epsilon_SEC}
\end{equation}
 
$R_\tinySEC(t)$ represents the global thermal resistance in the heat transfer process between the secondary and intermediate fluids, where once again the three terms shown in Equation Set \eqref{eq:R_SEC} are involved. Their expressions are detailed in Equation Set \eqref{eq:R_SEC_DESCRIPTION}.

\begin{equation}
	R_\tinySEC(t) \,=\, R_\tinySEC^\tinyCONVINT(t)  + R_\tinySEC^\tinyCONDWALL  + R_\tinySEC^\tinyCONVEXT(t) 
	\label{eq:R_SEC}
\end{equation}

\begin{equation}
	\begin{aligned}
	R_\tinySEC^\tinyCONDWALL \,&=\, \frac{\ln{\frac{r_{\!\tinySEC}+e_\tinySEC}{r_{\!\tinySEC}}}} {\kappa_\tinySEC^\tinyWALL\,2\pi\,l_\tinySEC} \\
	R_\tinySEC^\tinyCONVINT(t) \!&=\! \frac{1}{\alpha_\tinySEC^\tinyINT(t)\,2\pi\,r_{\!\tinySEC}\,l_\tinySEC}\; \\  R_\tinySEC^\tinyCONVEXT(t) \!&=\! \frac{1}{\alpha_\tinySEC^\tinyEXT(t)\,2\pi\,(r_{\!\tinySEC}\!\!+\!e_\tinySEC)\,l_\tinySEC}
	\end{aligned}
	\label{eq:R_SEC_DESCRIPTION}
\end{equation}

Eventually, the heat transfer coefficients are computed from the \emph{Nusselt} number, using the appropriate correlation. In the case of $\alpha_\tinySEC^\tinyEXT(t)$, \ref{secAppendixExternalConvectionPipes} describes such a correlation. In the case of $\alpha_\tinySEC^\tinyINT(t)$, the description is given in \ref{secAppendixInternalForcedConvectionPipes}.

$\dot Q_\tinySEC(t)$ is also related to the inlet-outlet temperature difference in the
secondary fluid pipes, through the heat capacity rate, as described in Equation \eqref{eq:Q_SEC_aux}, or alternatively through the mass average temperature inside the pipe and the pipe wall temperature, as indicated in Equation \eqref{eq:Q_SEC_aux2}.

\begin{equation}
	\dot Q_\tinySEC(t) \,=\, C_\tinySEC(t)\,(T_\tinySEC^\tinyOUT(t) - T_\tinySEC^\tinyIN(t)) 
	\label{eq:Q_SEC_aux}
\end{equation}

\begin{equation}
	\dot Q_\tinySEC(t) \,=\, \frac{T_\tinySEC^\tinyWALL(t) - \frac{1}{2}\left(T_\tinySEC^\tinyIN(t) + T_\tinySEC^\tinyOUT(t)\right)}{R_\tinySEC^\tinyCONVINT(t) + R_\tinySEC^\tinyCONDWALL}  
	\label{eq:Q_SEC_aux2}
\end{equation}

\subsection{State representation of the continuous model}

In the case of the continuous model, the state of the system will be represented by the vector indicated in Equation \eqref{eqVectorEstadoContinuo}. It includes the indication of either charging or discharging process taking place (remember that, in the simplified continuous model, only full charge or full discharge operations are possible, thereby a binary value suffices), then the radius of the freezing/melting boundary inside the PCM capsules, the radius of the PCM capsules, and finally the temperature of the intermediate fluid.

\begin{equation}
 	\boldsymbol{x}(t) \,=\, 
 	\begin{bmatrix} 
 	cd(t) \\ 
 	r(t) \\ 
 	r_\tinyPCM (t) \\ 
 	T_\tinyINT(t) \\
 	\end{bmatrix}
 	\label{eqVectorEstadoContinuo}
\end{equation}


\section{Detailed discrete model} \label{secCompleteDiscreteModel}

The main limitation of the continuous model described above is that only strict full
charging/discharging operations can be simulated: a new charging operation begins at the very moment when the PCM capsules are completely melted ($r(t=0)=r_\tinyPCM(t=0)=r_{\!\tinyPCM}^{\,\tinyMAX}$), and finishes when $r(t)=0$, that is just when the centre of the PCM capsule becomes frozen. Conversely, the discharging operation starts exactly at the point where the charging operation ended and finishes when the inward melting boundary reaches the centre of the sphere.

The discrete model to be introduced here is not restricted to full charging or discharging operations, but it can also represent the dynamic evolution of the system during any series of partial charging/discharging operations, in such a way that an arbitrary number of moving freezing/melting boundaries could be present at the same time inside the PCM capsules. In practice, this would mean a potentially infinite-dimensional state vector, since a single value of $r(t)$,
present in Equation \eqref{eqVectorEstadoContinuo}, is insufficient to represent the energy level of the capsule. This is where the discretised approach comes handy.

\subsection{Description of the detailed discrete model}

\begin{figure}[htbp]
    \centerline{\includegraphics[width=7cm,trim = 0 21 0 18,clip]
    {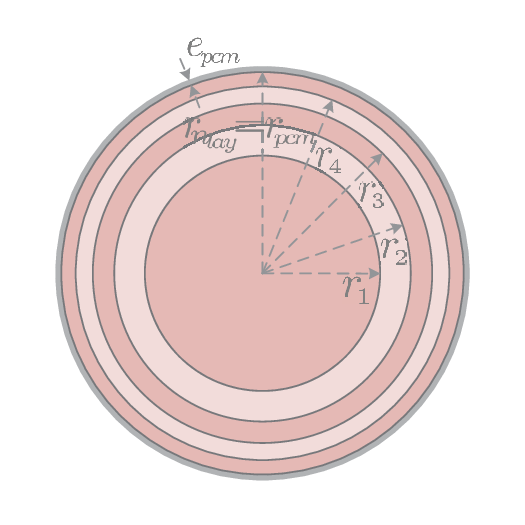}}
    \caption{Example scheme of the discretised PCM capsule for $n_\tinyLAY = 5$.}
    \label{figPCMcapsuleOnion}
\end{figure}

A discretisation of the continuous PCM-capsule volume is defined, according to a predefined number of layers, $n_\tinyLAY$, each of them including the exact same PCM mass $m_\tinyLAY$. The discretised PCM capsule is represented in Figure \ref{figPCMcapsuleOnion} for $n_\tinyLAY=5$.

In the case of the discrete model, the state vector has the structure indicated in Equation \eqref{eq:x}, where the state of each layer $k$ is represented by its specific enthalpy, $h_\tinyPCMK(t)$, together with the temperature of the intermediate fluid $T_\tinyINT(t)$.

\begin{equation}
    \boldsymbol{x}(t) \,=\, \begin{bmatrix} h_\tinyPCMOne(t) \\ h_\tinyPCMTwo(t) \\ \cdots \\ h_\tinyPCMNLAY(t) \\ T_\tinyINT(t) \end{bmatrix}
    \label{eq:x}
\end{equation}

Regarding the system modelling, the only difference with respect to the previously described model is that the term $\dot Q_\tinyPCM(t)$ in Equation (\ref{eqPrincipal}) is now obtained in a different way. Energy balance in each particular layer $k$ is ruled by Equation Set \eqref{eq:hdot}, for $k=1, 2, \,\ldots ,\,n_\tinyLAY$.

\begin{equation}
   \begin{aligned}
   \dfrac{d h_\tinyPCMK(t)}{dt} \,&=\, \frac{1}{\rho_\tinyPCMK(t)\, V_\tinyPCMK(t)} \left( \dot Q_\tinyPCMK^{\,\tinyEXT}(t) - \dot Q_\tinyPCMK^{\,\tinyINT}(t) \right) \\
   \dot Q_\tinyPCMK^{\,\tinyEXT} \,&=\, \frac{T_\tinyPCMKMas(t) - T_\tinyPCMK(t)}{R_\tinyPCMKMas^\tinyCOND(t)} \\
   \dot Q_\tinyPCMK^{\,\tinyINT} \,&=\, \frac{T_\tinyPCMK(t) - T_\tinyPCMKMenos(t)}{R_\tinyPCMK^\tinyCOND(t)} \\
   \end{aligned}
   \label{eq:hdot}
\end{equation}

In Equation Set \eqref{eq:hdot} $V_\tinyPCMK (t)$ is the volume of every individual layer ($V_\tinyPCMK(t) = \frac{m_\tinyLAY}{\rho_\tinyPCMK(t)}$), while $\dot Q_\tinyPCMK^{\,\tinyEXT}(t)$ is the cooling power transferred to the neighbour outer layer (layer number $k\!+\!1$) and $\dot Q_\tinyPCMK^{\,\tinyINT}(t)$ is the cooling power received from the inner neighbour layer (layer number $k\!-\!1$). The inner-most layer verifies $\dot Q_\tinyPCMOne^{\,\tinyINT}(t)\!=\! 0$, while the
outermost layer is that one ultimately exchanging cooling power with the intermediate fluid, in such a way that $\dot Q_\tinyPCMNLAY^{\,\tinyEXT}(t)\!=\! \dot Q_\tinyPCM(t)$.

For each layer $k$, the dependence between its temperature and its specific enthalpy has three alternatives, according to Figure \ref{figThDiagrams}, as indicated in Equation Set \eqref{eq:h_T}.

\begin{equation}
   \begin{aligned}
        T_\tinyPCMK(t) &= T_\tinyPCM^\tinyLAT & \qquad & \text{if} \quad &h_\tinyPCM^\tinyLATMIN \le &h_\tinyPCMK(t) \le &h_\tinyPCM^\tinyLATMAX \\
        T_\tinyPCMK(t) &= T_\tinyPCM^\tinyLAT - \frac{h_\tinyPCM^\tinyLATMIN - h_\tinyPCMK(t)}{c_{\!p_\tinyPCMSOL}} & \qquad & \text{if} \quad  &h_\tinyPCM^\tinyLATMIN > & h_\tinyPCMK(t) & \\
        T_\tinyPCMK(t) &= T_\tinyPCM^\tinyLAT + \frac{h_\tinyPCMK(t) - h_\tinyPCM^\tinyLATMAX}{c_{\!p_\tinyPCMLIQ}} & \qquad & \text{if} \quad  & & h_\tinyPCMK(t) > &h_\tinyPCM^\tinyLATMAX
    \end{aligned}
    \label{eq:h_T}
\end{equation}

The \emph{charge ratio} of each layer $k$, $\gamma_\tinyPCMK(t)$, can be defined according to the Equation \eqref{eq:charge_ratio_k}. It allows to weight the thermodynamic properties of each layer $k$, such as density and thermal conductivity (required when computing the thermal resistance $R_\tinyPCMK^\tinyCOND$), according to the mentioned \emph{charge ratio}, as indicated in Equation Sets \eqref{eq:rho_h} and \eqref{eq:k_h}.

\begin{equation}
	\gamma_\tinyPCMK(t) = \frac{h_\tinyPCM^\tinyLATMAX - h_\tinyPCMK(t)}{h_\tinyPCM^\tinyLATMAX - h_\tinyPCM^\tinyLATMIN} \\
	\label{eq:charge_ratio_k}
\end{equation}

\begin{equation}
	\begin{aligned}
	\rho_\tinyPCMK(t) &= \rho_\tinyPCM^\tinyLATMAX + \gamma_\tinyPCMK(t)\,(\rho_\tinyPCM^\tinyLATMIN - \rho_\tinyPCM^\tinyLATMAX) & \;\;\;\;\;\;\;\;\;\; & \text{if} \quad &h_\tinyPCM^\tinyLATMIN \le &h_\tinyPCMK(t) \le &h_\tinyPCM^\tinyLATMAX \\
	\rho_\tinyPCMK(t) &= \rho_\tinyPCM^\tinyLATMIN & \;\;\;\;\;\;\;\;\;\; & \text{if} \quad  &h_\tinyPCM^\tinyLATMIN > & h_\tinyPCMK(t) & \\
	\rho_\tinyPCMK(t) &= \rho_\tinyPCM^\tinyLATMAX & \;\;\;\;\;\;\;\;\;\; & \text{if} \quad  & & 	h_\tinyPCMK(t) > &h_\tinyPCM^\tinyLATMAX
	\end{aligned}
	\label{eq:rho_h}
\end{equation}

\begin{equation}
	\begin{aligned}
	\kappa_\tinyPCMK(t) &= \kappa_{\tinyPCM,\,\tinyEFF}^\tinyLATMAX + \gamma_\tinyPCMK(t)\,(\kappa_\tinyPCM^\tinyLATMIN - \kappa_{\tinyPCM,\,\tinyEFF}^\tinyLATMAX) & \; & \text{if} \;\; &h_\tinyPCM^\tinyLATMIN \le &h_\tinyPCMK(t) \le &h_\tinyPCM^\tinyLATMAX \\
	\kappa_\tinyPCMK(t) &= \kappa_\tinyPCM^\tinyLATMIN & \; & \text{if} \;\;  &h_\tinyPCM^\tinyLATMIN > & h_\tinyPCMK(t) & \\
	\kappa_\tinyPCMK(t) &= \kappa_{\tinyPCM,\,\tinyEFF}^\tinyLATMAX & \; & \text{if} \;\;  & & 	h_\tinyPCMK(t) > &h_\tinyPCM^\tinyLATMAX
	\end{aligned}
	\label{eq:k_h}
\end{equation}

Moreover, it is possible to compute the external radius $r_\tinyPCMK(t)$ of every layer as shown in Equation \eqref{eq:radius_k}, being $r_{\tinyPCM,0} = 0$. Note that $r_{\tinyPCM,n_\tinyLAY}(t)$ corresponds to the state variable $r_\tinyPCM(t)$ of the continuous model. 

\begin{equation}
	r_\tinyPCMK(t) = \left( {r_\tinyPCMKMenos(t)}^3 + \frac{3\,V_\tinyPCMK(t)}{4 \, \pi} \right)^{1/3}\\
	\label{eq:radius_k}
\end{equation}

\subsection{Comparison continuous versus discrete models}

In order to verify the validity of the discrete model, comparative simulations regarding exclusive charging and discharging processes have been carried out.

\begin{figure}[h]
    \centerline{\includegraphics[width=9cm,angle=0]{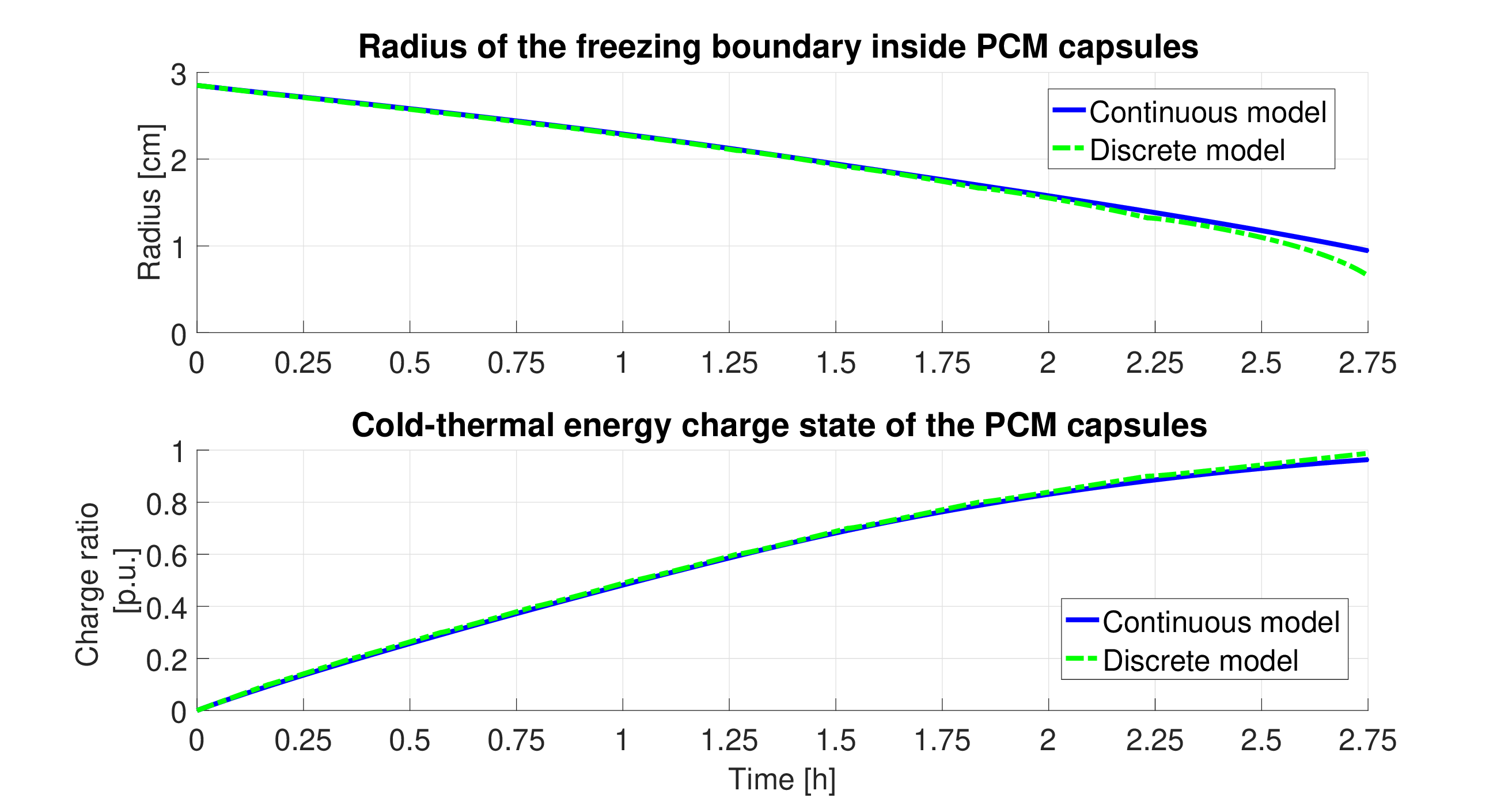}}
    \caption{Comparison of the radius (upper plot) and \emph{charge ratio} (lower plot) during a charging operation between continuous and discrete models.}
    \label{figComparisonContDiscr_PCMRadius_charge}
\end{figure}

A typical full charging operation in latent zone is simulated, starting from a discharged initial state. Figure \ref{figComparisonContDiscr_PCMRadius_charge} shows the comparison of the radius of the inward freezing boundary inside each PCM capsule, along with the \emph{charge ratio} of the PCM capsule. The discrete model is working with ten layers ($n_\tinyLAY=10$).

\begin{figure*}[h]
    \centerline{\includegraphics[width=15cm,angle=0]{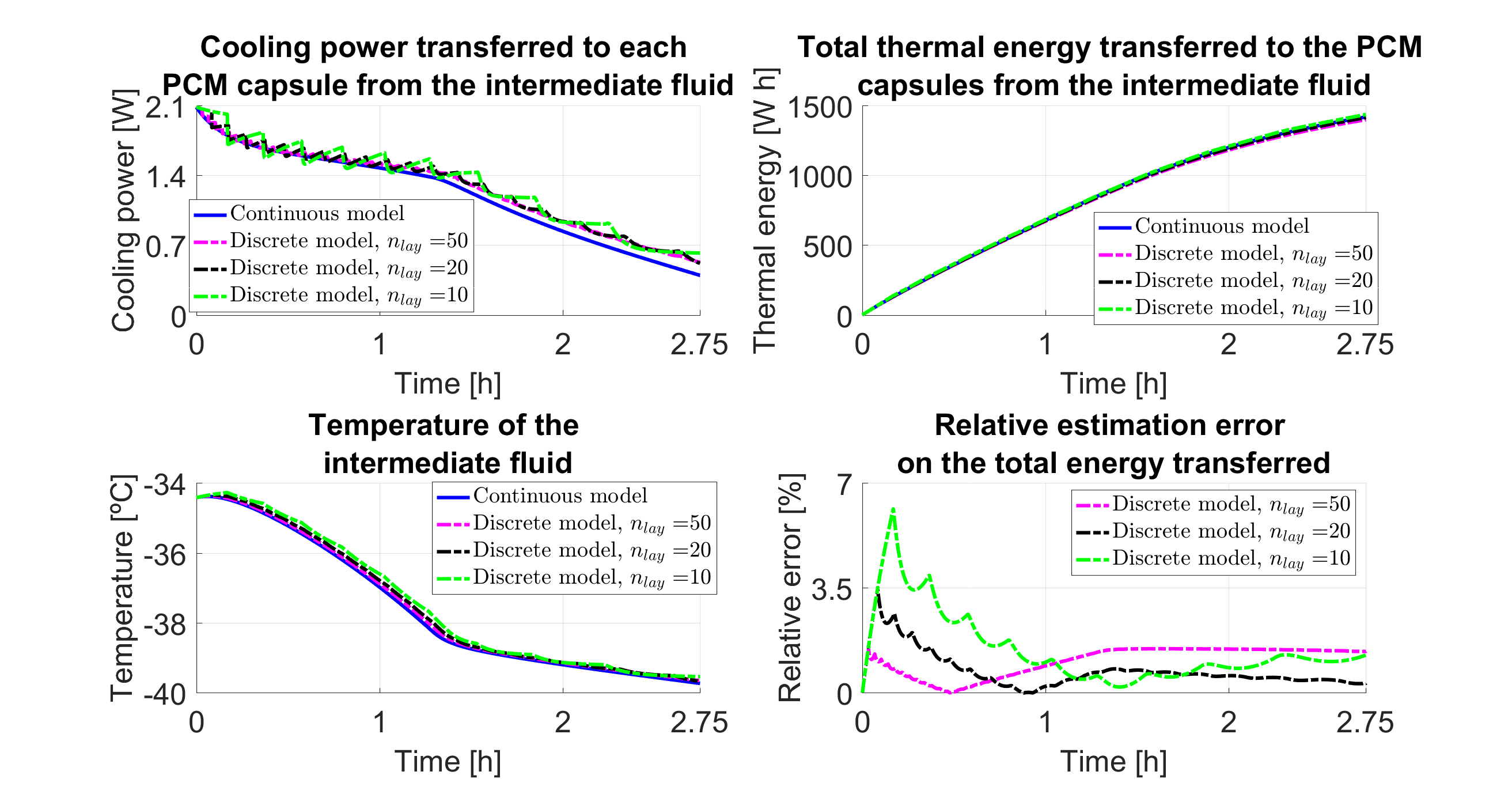}}
    \caption{Comparison of the transferred cooling power (upper left subplot) between each PCM capsule and the intermediate fluid; the temperature of the intermediate fluid (lower left subplot); and the total thermal energy (right subplots) transferred between the PCM and intermediate fluid during the charging operation.}
    \label{figComparisonContDiscrVarios_PotenciaEnergiaTotalPCMError_charge}
\end{figure*}

Figure \ref{figComparisonContDiscrVarios_PotenciaEnergiaTotalPCMError_charge} represents the comparative evolution of the cooling power transferred to each PCM capsule from the intermediate fluid, $-\dot Q_\tinyPCM(t)$, during this charging process; the temperature of the intermediate fluid; the total amount of energy (in absolute value) transferred, considering the $n_\tinyPCM$ PCM capsules inside the tank; and eventually the corresponding relative error of the discrete model with respect to the continuous one. Please note that the total energy considered corresponds to the sum of the latent energy stored in the capsule core and also the sensible energy stored in the frozen shell, as described in the work by Bédécarrats \emph{et al.} \cite{bedecarrats2009bstudy}. In this case, the comparison is extended to $n_\tinyLAY=\{10,20,50\}$.

Some oscillations are noticed in the upper left subplot of Figure \ref{figComparisonContDiscrVarios_PotenciaEnergiaTotalPCMError_charge}. Concerning the discrete model, the cooling power transferred between each PCM capsule and the intermediate fluid is expected to vary sharply when the specific enthalpy of a certain spherical layer comes out of the latent zone, which means that the global thermal resistance is increased by the thermal resistance related to heat conduction of the corresponding layer. It causes the cooling power to be sharply reduced. However, while a certain layer remains in latent zone, the cooling power also evolves according to the temperature of the intermediate fluid, shown in the lower left subplot of Figure \ref{figComparisonContDiscrVarios_PotenciaEnergiaTotalPCMError_charge}. This is the reason why some oscillations appear when analysing the cooling power during a charging or discharging operation. These oscillations are indeed smaller as the number of layers increases, since the additional thermal resistance due to each layer is smaller. Moreover, the thermal conductivity depends on the specific enthalpy of the corresponding layer, as shown in Equation Set \eqref{eq:k_h} of the revised manuscript, in such a way that it is lower when the \emph{charge ratio} of each PCM layer $\gamma_\tinyPCMK$ is close to zero. It causes the thermal resistance corresponding to heat conduction to increase and it reduces the cooling power.

As it can be seen in the lower right subplot of Figure \ref{figComparisonContDiscrVarios_PotenciaEnergiaTotalPCMError_charge}, the relative error when comparing the discrete models and the continuous model regarding the total thermal energy transferred is under 7\%. Moreover, these simulations show that not much accuracy is gained by increasing the number of layers regarding the total energy transferred. Note the satisfactory accuracy of the discrete model, achieved using a relatively low number of layers: $n_\tinyLAY$ = 10. Similar qualitative results have been obtained when studying a full discharging operation, starting from a completely charged initial state. For the sake of brevity, the simulation results are not shown here.

It is concluded that just 10 layers can lead to an accurate approximation of the continuous model of the PCM-capsule behaviour, combined with an affordable computational cost. This is easily understandable if we bring to mind the relatively small encapsulation size chosen for the PCM capsules: $r_{\!\tinyPCM}^{\,\tinyMAX} <$ 30 mm.

\subsection{Series of partial charging/discharging operations}

It has been previously stated that the main limitation of the continuous model described in Section \ref{secSystemModelling} is that only strict full charging/discharging operations can be simulated. However, the operation of the PCM-based TES combined with the refrigeration system is more likely to be subjected to partial charging/discharging operations, according to the scheduling strategy defined by the control system. Therefore, it seems interesting to study the results of the discrete model in such scenarios, where an arbitrary number of moving freezing/melting boundaries could be present at the same time inside the PCM capsules, as shown later.

\begin{figure}[htbp]
    \centerline{\includegraphics[width=9cm,angle=0]{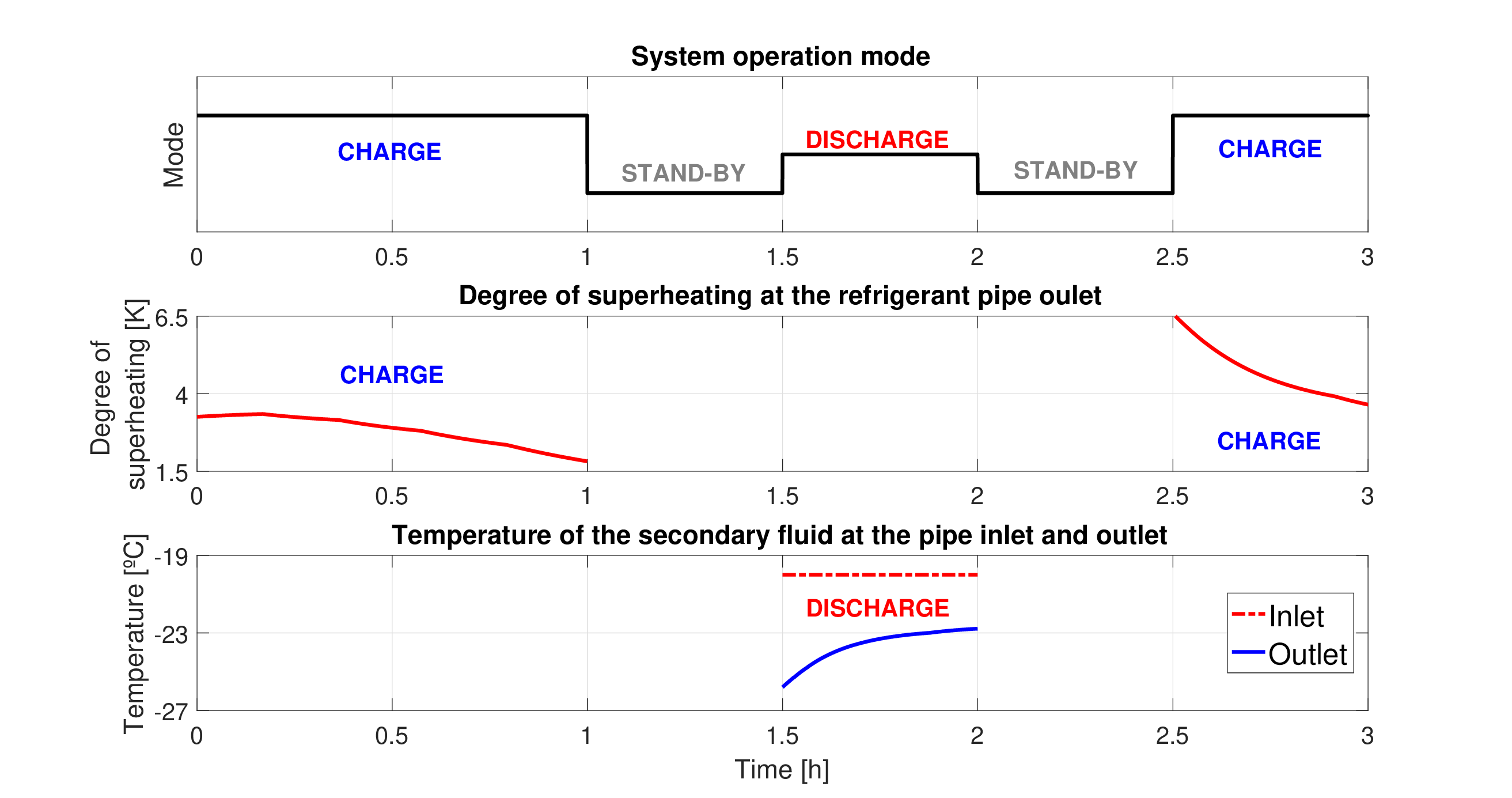}}
    \caption{Series of partial charging/discharging operations, along with the variables related to the refrigerant and the secondary fluid.}
    \label{fig_Refr_Sec_Mode}
\end{figure}

In this subsection a more complex simulation scenario, beyond full charging/discharging processes, is assumed. A series of partial charging/discharging operations with a short stand-by period is simulated. The PCM capsules are initially completely discharged. The particular sequence is as follows: 1 hour charge, 0.5 hours stand-by, 0.5 hours discharge, 0.5 hours stand-by and 0.5 hours charge. This sequence is also represented in Figure \ref{fig_Refr_Sec_Mode}, where some variables
related to the refrigerant and the secondary fluid are also depicted.

It is observed in Figure \ref{fig_Refr_Sec_Mode} that, during the charging processes, the refrigerant comes out of the pipe as superheated vapour, with positive degree of superheating. During these processes, the temperatures corresponding to the secondary fluid are not shown in the lowest subplot of Figure \ref{fig_Refr_Sec_Mode}, since no secondary fluid circulates. During the stand-by periods all variables related to the refrigerant and secondary fluid are again not shown, since none of them circulates. However, during the discharge, the temperature of the secondary fluid at the pipe outlet shows to be lower than that at the pipe inlet, which implies that the secondary fluid cools while discharging the TES. The variable related to the refrigerant is not shown in the central subplot of Figure \ref{fig_Refr_Sec_Mode}, since no refrigerant circulates during this period.

\begin{figure}[htbp]
    \centerline{\includegraphics[width=9cm,angle=0]{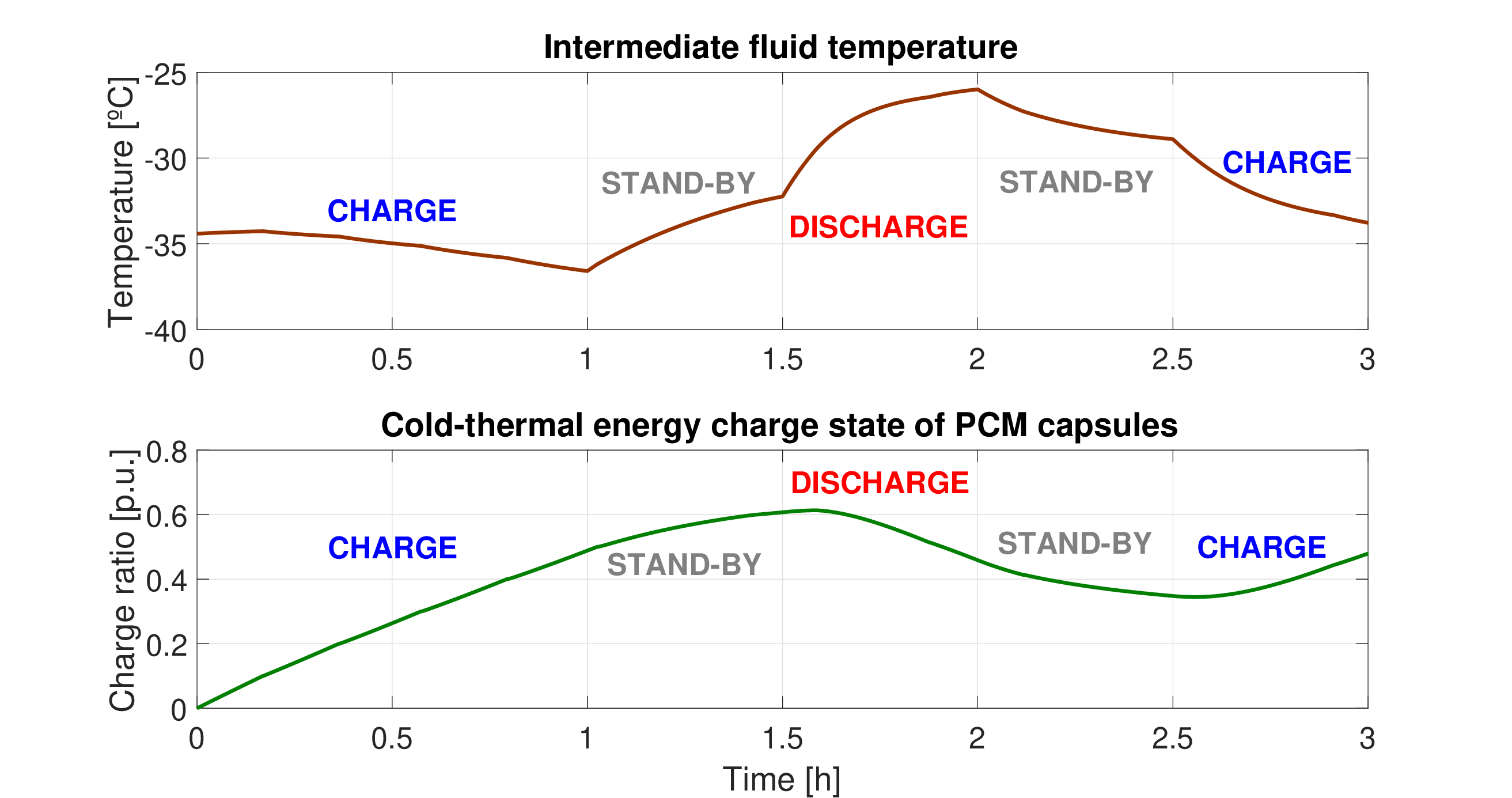}}
    \caption{Temperature of the intermediate fluid and \emph{charge ratio} of the PCM capsules.}
    \label{fig_Tint_gamma}
\end{figure}

Figure \ref{fig_Tint_gamma} shows the temperature of the intermediate fluid and the \emph{charge ratio} of the PCM capsules. Note that, after the first charging process finishes and the first stand-by period begins, the \emph{charge ratio} still increases, because the temperature of the intermediate fluid is still lower than the PCM melting point, which means that the PCM capsules continue to store cold-energy. This situation would persist until the intermediate fluid achieved the melting point, but it is disrupted by the discharge. During the first minutes of the discharging process, the temperature of the intermediate fluid is still lower than the PCM melting point, thus the PCM capsules continue to store cold-energy, but due to the heat transfer to the secondary fluid, the temperature of the intermediate fluid exceeds quickly the melting point and the PCM capsules release the cold-energy previously stored.

Similarly, after the discharging process finishes and the second stand-by period begins, the \emph{charge ratio} still decreases, because the temperature of the intermediate fluid is still greater than the PCM melting point, in such a way that the PCM capsules continue to release cold-energy. Eventually, during the first minutes of the second charge, the temperature of the intermediate fluid is still
greater than the PCM melting point, thus the PCM capsules continue to release cold-energy, but due to the heat transfer to the refrigerant, the temperature of the intermediate fluid comes down quickly below the melting point and the PCM capsules begin again to store cold-energy.

\begin{figure}[htbp]
    \centerline{\includegraphics[width=9cm,angle=0]{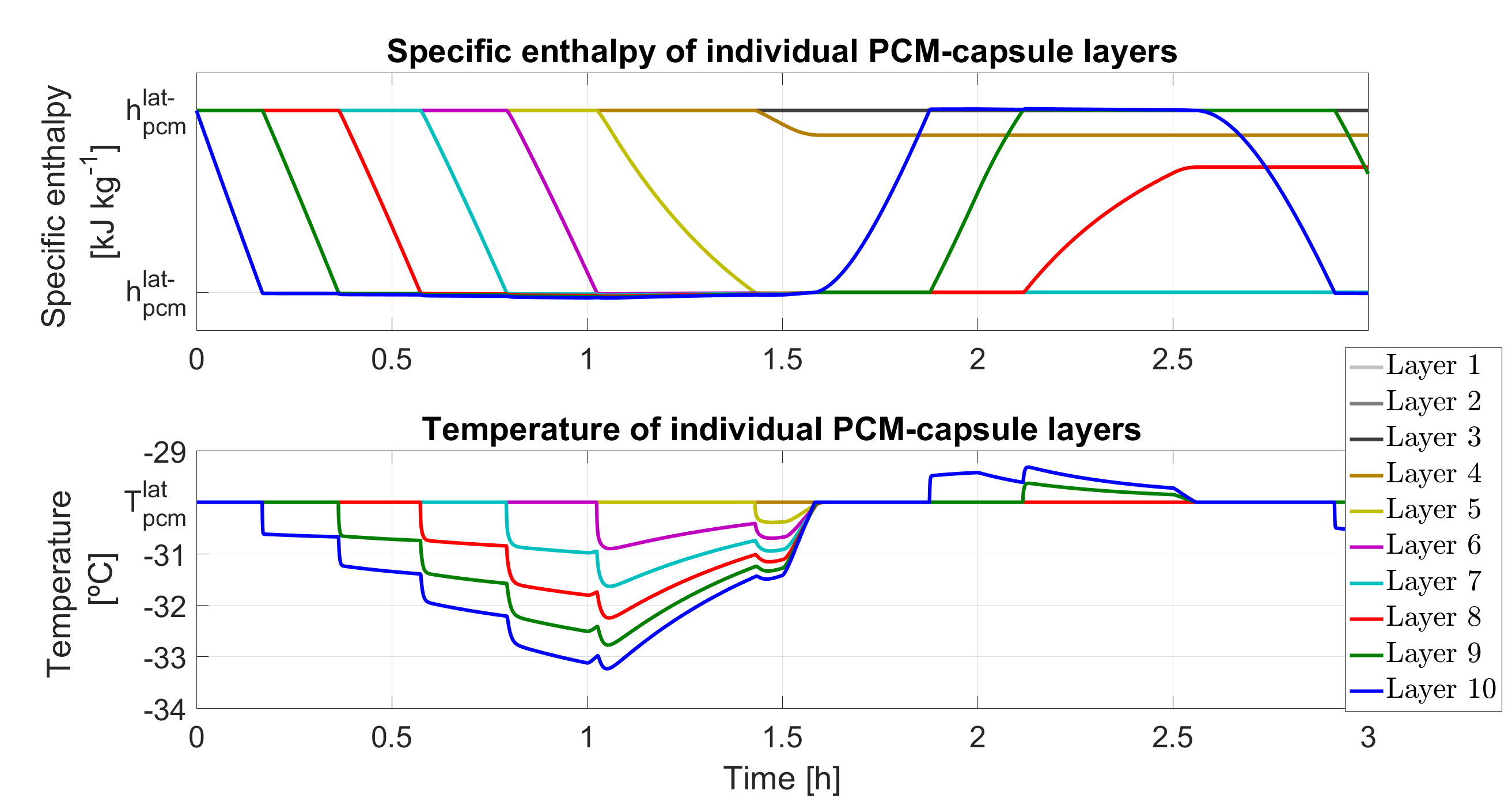}}
    \caption{Evolution of specific enthalpy (upper plot) and temperature (lower plot) of the individual PCM-capsule layers during a series of charging/discharging operations.}
    \label{fig_h_T_capas}
\end{figure}

Figure \ref{fig_h_T_capas} is very interesting, where the evolution of the specific enthalpy and temperature of the individual PCM-capsule layers is shown. It can be observed that, during the first one-hour charging stage, there is time for the four most external layers to reach minimum enthalpy levels, while the charge finishes when layer 6 is still in latent zone. During the next half-hour stand-by period, however, due to thermal inertia, layer 6 keeps on lowering its enthalpy level, layer 5 changes phase completely and even layer 4 decreases its specific enthalpy below $h_{\!\tinyPCM}^{\tinyLATMAX}$ before this evolution being disrupted by the beginning of the discharge. During the half-hour discharging operation there is only time for the most external layer to completely melt, but it is worth noticing that layer 4 keeps its intermediate latent state, since the temperature of its adjacent layers corresponds to the melting point and there is thermal equilibrium between them. During the next half-hour stand-by period, layer 9 completely melts and there is time for the layer 8 to gain enthalpy due to the thermal inertia of the intermediate fluid, being truncated by the final charging operation. Note also that layer 8 keeps its intermediate latent state, since, until layer 9 freezes completely, there is no heat transfer mechanism for layer 8 to modify its specific enthalpy.

\begin{figure*}[h]
    \centering
    \subfigure[$t$ = 2.25 h]{
        \includegraphics[width=6.5cm,angle=0]{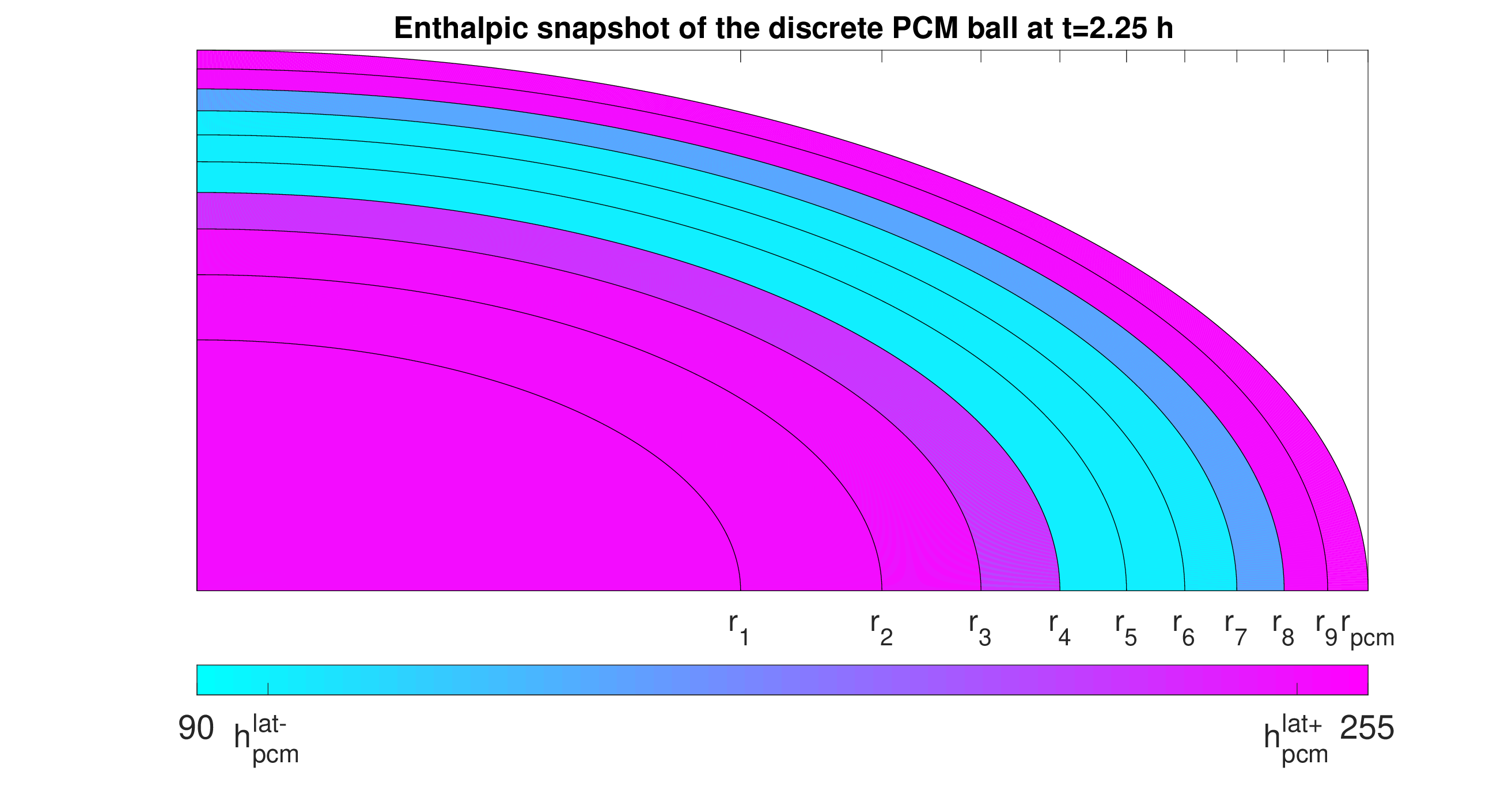}
        \label{fig_fotos_bolas_1}
    }\subfigure[$t$ = 3 h]{
        \includegraphics[width=6.5cm,angle=0]{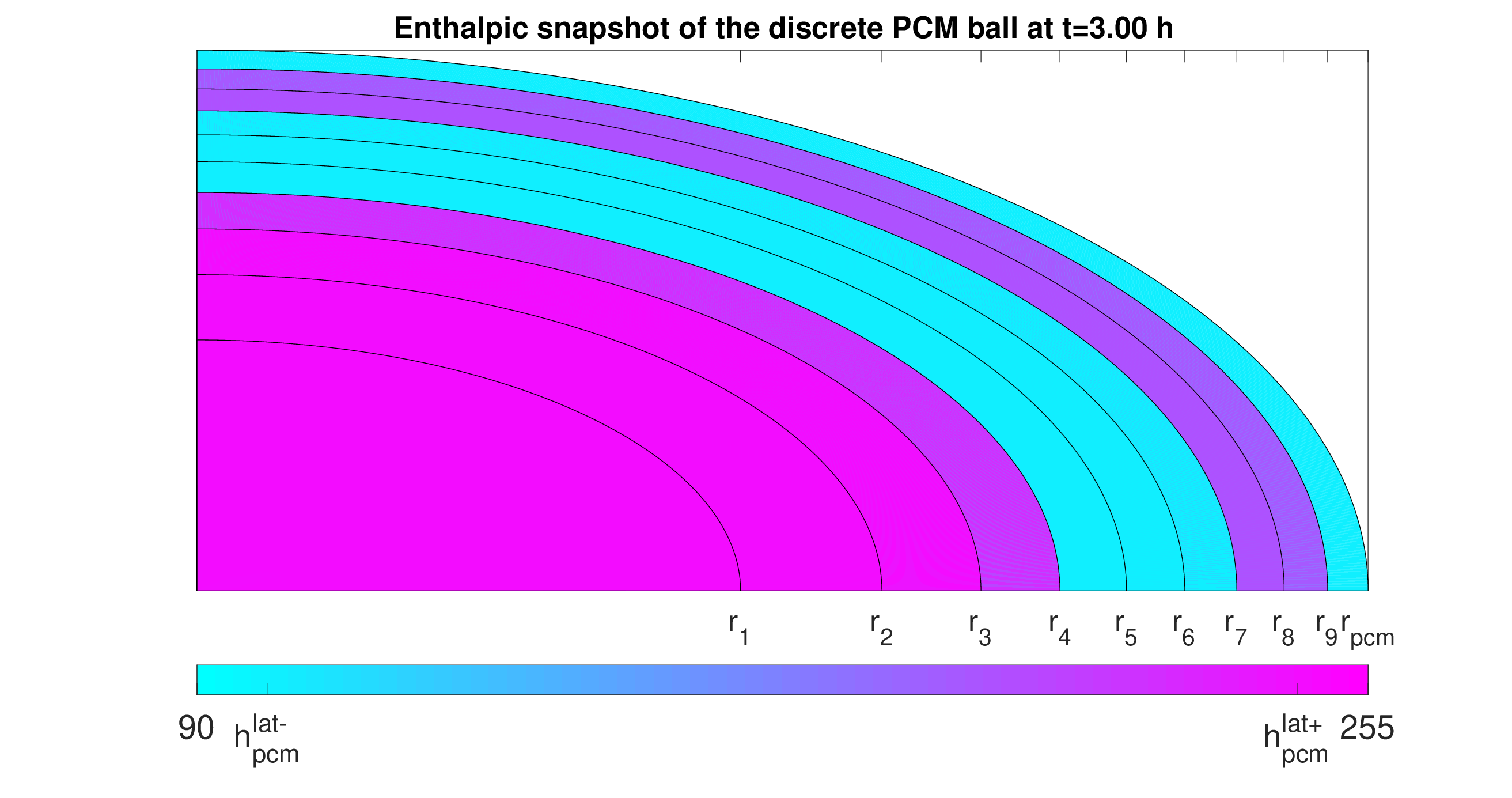}
        \label{fig_fotos_bolas_2}
    }
    \caption{PCM-capsule \emph{enthalpic snapshots} at two instants highlighting the energy state of all layers.}
    \label{fig_fotos_bolas}
\end{figure*}

Figure \ref{fig_fotos_bolas} shows \emph{enthalpic snapshots} of the PCM capsule at two instants during the simulation: at $t\!=\!$ 2.25 h (left subplot, during the discharging process) and at $t\!=\!$ 3 h (right subplot, at the end of the second charging process), where the energy state of all PCM-capsule layers can be observed and the \emph{stuck} layers 4 and 8 are highlighted.

These simulations prove the need for the discrete-spatial model, since the continuous one is not able to provide these results.


\section{Conclusions and future work} \label{secConclusions}

The design of a novel PCM-based cold-energy storage system for a laboratory refrigeration plant has been tackled in this paper. Given a series of design specifications, the encapsulated PCM and intermediate fluid have been chosen for efficient heat transfer.

Firstly, a dynamic continuous model only valid for full charging/discharging cycles has been developed. This model has been used for the design of specific parameters, such as the TES system volume, heat transfer area, number and volume of the PCM capsules, etc. Moreover, this model has provided other key data, such as the needed time for a full charging/discharging cycle, or the evolution of the \emph{charge ratio} along time.

Secondly, using the continuous model as a keystone, a dynamic discrete-spatial model has been also developed, taking into account the fact that the continuous model cannot be used for partial charging/discharging operations. Some simulations comparing the results obtained with both models in a charging cycle have been provided, where it is concluded that only ten spherical layers are enough to describe properly the dominating system dynamics. This number of layers has been selected as a trade-off between small errors in the comparison of the results of both models, and low computational load.

Additionally, some partial charging/stand-by/discharging simulations have been performed using the discrete model, providing coherent results and proving the need  for the discrete model when analysing more complex charging/discharging scenarios.


\section*{Acknowledgements}
The authors would like to acknowledge MCeI (Grants DPI2015-70973-R and DPI2013-44135-R) for funding this work.


\appendix

\section{Correlations used} \label{appendixCorrelations}

This Appendix lists a series of correlations used to compute the heat transfer coefficients from the \emph{Nusselt} number.

\subsection{Correlation for external natural convection in spheres} \label{secAppendixExternalConvectionSpheres}

In the case of the PCM capsules, the heat transfer coefficient, $\alpha_\tinyPCM^\tinyEXT(t)$, is computed from the \emph{Nusselt} number, as represented in Equation \eqref{eqn:alpha_spheres}, using the Churchill correlation for external natural convection in spheres \cite{bergman2011fundamentals}, as shown in Equation \eqref{eqn:Nu_spheres}.

\begin{equation}
	\alpha_\tinyPCM^\tinyEXT(t)=\frac{\kappa_\tinyINT}{2\,(r_{\!\tinyPCM}\!\!+\!e_\tinyPCM)}\Nu_\tinyPCM^\tinyEXT(t)
	\label{eqn:alpha_spheres}
\end{equation}

\begin{equation}
	\Nu_\tinyPCM^\tinyEXT(t)=2+\frac{0.589 \,\Ra_\tinyPCM^\tinyEXT(t)^{1/4}}{\left[1 +
	(0.469/\Pr_\tinyPCM^\tinyEXT)^{9/16}\right]^{4/9}} 
	\label{eqn:Nu_spheres}
\end{equation}

On the one hand, the \emph{Rayleigh} number is obtained using Equation \eqref{eqn:Ra_spheres}. The \emph{Grashof} and \emph{Prandtl} numbers, on the other hand, are computed as shown in Equations \eqref{eqn:Gr_spheres} and \eqref{eqn:Pr_spheres}.

\begin{equation}
	\Ra_\tinyPCM^\tinyEXT(t)=\Gr_\tinyPCM^\tinyEXT\!(t)\Pr_\tinyPCM^\tinyEXT 
	\label{eqn:Ra_spheres}
\end{equation}

\begin{equation}
	\Gr_\tinyPCM^\tinyEXT(t)=\frac{g\,\beta_\tinyINT\,|T_\tinyPCM^\tinyWALL(t)-T_\tinyINT(t)|\,
	(2\,(r_{\!\tinyPCM}\!\!+\!e_\tinyPCM))^3}{\left(\mu_\tinyINT/\rho_\tinyINT\right)^2}
	\label{eqn:Gr_spheres}
\end{equation}

\begin{equation}
	\Pr_\tinyPCM^\tinyEXT=\frac{c_{\!p_{\,\tinyINT}}\mu_\tinyINT}{\kappa_\tinyINT}
	\label{eqn:Pr_spheres}
\end{equation}

Actually, $\rho_\tinyINT,\,\beta_\tinyINT,\,\kappa_\tinyINT,\,c_{\!p_\tinyINT},\,$ and $\mu_\tinyINT$ are not constants, but computed as functions of the average film temperature, $\frac{1}{2}(T_\tinyPCM^\tinyWALL(t) + T_\tinyINT(t))$, and the pressure of the intermediate fluid, which is assumed to work at constant atmospheric pressure. These thermodynamic functions are provided by the \emph{CoolProp} tool \cite{CoolProp}.

\subsection{Correlation for external natural convection in tubular pipes} \label{secAppendixExternalConvectionPipes}

In our system, this type of correlation is used to compute the external heat transfer coefficients of the secondary fluid and the refrigerant pipes: $\alpha_\tinySEC^\tinyEXT(t),\, \alpha_\tinyREFTWO^\tinyEXT(t),\,$ and $\alpha_\tinyREFVAP^\tinyEXT(t)$. In the case of the refrigerant, we need to separately compute the heat transfer coefficient for the two-phase zone and the superheated vapour zone.

First, in the case of the secondary fluid, the convective coefficient is computed using Equation \eqref{eqn:alpha_ext_sec}, while the \emph{Nusselt} number is calculated using Equation \eqref{eqn:Nu_ext_sec}.

\begin{equation}
	\alpha_\tinySEC^\tinyEXT(t)=\frac{\kappa_\tinyINT}{l_{\!\tinySEC}} \,\Nu_\tinySEC^\tinyEXT(t)
	\label{eqn:alpha_ext_sec}
\end{equation}

\begin{equation}
	\left\{ \begin{matrix}
	& \Nu_\tinySEC^\tinyEXT(t) = 0.59\, \Ra_\tinySEC^\tinyEXT(t)^{1/4} \,\quad\qquad &\Ra_\tinySEC^\tinyEXT(t) \in[10^4,\,10^9]\\ \\
	& \Nu_\tinySEC^\tinyEXT(t) = 0.10\, \Ra_\tinySEC^\tinyEXT(t)^{1/3} \,\quad\qquad
	&\Ra_\tinySEC^\tinyEXT(t) \in[10^9,\,10^{13}]
	\end{matrix}\right.
	\label{eqn:Nu_ext_sec}
\end{equation}

This is the McAdams correlation \cite{bergman2011fundamentals}, and the distinction based on the \emph{Rayleigh} number depends mainly on the buoyancy-driven flow in the intermediate fluid; the first case corresponds to laminar flow, while in the second case turbulent flow is produced. The \emph{Rayleigh} number is obtained from Equation \eqref{eqn:Ra_ext_sec} as a function of the \emph{Grashof} and \emph{Prandtl} numbers, calculated as indicated in Equations \eqref{eqn:Gr_ext_sec} and \eqref{eqn:Pr_ext_sec}.

\begin{equation}
	\Ra_\tinySEC^\tinyEXT(t)=\Gr_\tinySEC^\tinyEXT\!(t)\Pr_\tinySEC^\tinyEXT\
	\label{eqn:Ra_ext_sec}
\end{equation}

\begin{equation}
	\Gr_\tinySEC^\tinyEXT(t)= \frac{g\,\beta_\tinyINT\,|T_\tinySEC^\tinyWALL(t)-T_\tinyINT(t)|\,l_{\!\tinySEC}^3}{\left(\mu_\tinyINT/\rho_\tinyINT\right)^2}
	\label{eqn:Gr_ext_sec}
\end{equation}

\begin{equation}
	\Pr_\tinySEC^\tinyEXT \,=\, \frac{c_{\!p_{\,\tinyINT}}\mu_\tinyINT}{\kappa_\tinyINT}
	\label{eqn:Pr_ext_sec}
\end{equation}

Again, $\rho_\tinyINT,\,\beta_\tinyINT,\,\kappa_\tinyINT,\,c_{\!p_\tinyINT},\,$ and $\mu_\tinyINT$ are not constant, but computed as functions of the average film temperature, $\frac{1}{2}(T_\tinySEC^\tinyWALL(t) + T_\tinyINT(t))$, and the pressure of the intermediate fluid, which is assumed to work at constant atmospheric pressure.

In the case of the refrigerant in the two-phase zone, the previous expressions hold, replacing $l_\tinySEC$ and $T_\tinySEC^\tinyWALL(t)$ by $\zeta_\tinyREF(t)\,l_\tinyREF$ and $T_\tinyREFTWO^\tinyWALL(t)$, respectively. Moreover, in this case, $\rho_\tinyINT,\,\beta_\tinyINT,\,\kappa_\tinyINT,\,c_{\!p_\tinyINT},\,$ and $\mu_\tinyINT$ must be computed as functions of the corresponding average film temperature, $\frac{1}{2}(T_\tinyREFTWO^\tinyWALL(t) + T_\tinyINT(t))$, and the pressure of the intermediate fluid.

Similarly, for the superheated vapour zone, $l_\tinySEC$ and $T_\tinySEC^\tinyWALL(t)$ in the previous expressions must be replaced by $(1-\zeta_\tinyREF(t))l_\tinyREF$ and $T_\tinyREFVAP^\tinyWALL(t)$, respectively. Again, $\rho_\tinyINT,\,\beta_\tinyINT,\,\kappa_\tinyINT,\,c_{\!p_\tinyINT},\,$ and $\mu_\tinyINT$ must be computed as functions of the corresponding average film temperature, $\frac{1}{2}(T_\tinyREFVAP^\tinyWALL(t) + T_\tinyINT(t))$, and the pressure of the intermediate fluid (atmospheric pressure).

\subsection{Correlation for internal forced convection in tubular pipes} \label{secAppendixInternalForcedConvectionPipes}

This correlation is used with minor differences for both the internal forced convection of the secondary fluid and the internal forced convection of the refrigerant in the superheated vapour zone, $\alpha_\tinySEC^\tinyINT(t)$ and $\alpha_\tinyREFVAP^\tinyINT(t)$, respectively.

Regarding the secondary fluid, the convective coefficient can be computed using Equation \eqref{eqn:alpha_int_sec}.

\begin{equation}
	\alpha_\tinySEC^\tinyINT(t)=\frac{\kappa_\tinySEC}{2\,r_{\!\tinySEC}}\,\Nu_\tinySEC^\tinyINT(t)
	\label{eqn:alpha_int_sec}
\end{equation}

The particular correlation depends on the instantaneous value of the \emph{Reynolds} number: it is assumed that laminar flow occurs when $\Re_\tinySEC^\tinyINT(t) \le 3000$, whereas over this value, the flow is considered to be turbulent. Therefore, the conditional Equation Set \eqref{eqn:cond} has been considered, where $\psi$ is the friction coefficient of turbulent fluid inside smooth pipes, computed using Equation \eqref{eqn:f}.

\scriptsize
\begin{equation}
	\left\{ 
	\begin{matrix} 
	\Nu_\tinySEC^\tinyINT(t) &=& \dfrac{\dfrac{3.66}{tanh[2.264\Gz_\tinySEC^\tinyINT(t)^{-1/3}+1.7\Gz_\tinySEC^\tinyINT(t)^{-2/3}]} \,+0.0499\Gz_\tinySEC^\tinyINT(t)tanh(\Gz_\tinySEC^\tinyINT(t)^{-1})}{tanh(2.432(\Pr_\tinySEC^\tinyINT)^{1/6}\Gz_\tinySEC^\tinyINT(t)^{-1/6})}\,\quad\qquad
	&\Re_\tinySEC^\tinyINT(t) \le 3000
	\\ \\
	\Nu_\tinySEC^\tinyINT(t) &=& \dfrac{\psi\!/\!8 (\Re_\tinySEC^\tinyINT(t)-1000)
	\Pr_\tinySEC^\tinyINT(t)}{1+12.7\,(\psi\!/\!8)^{1/2}(\Pr_\tinySEC^\tinyINT(t)^{2/3}-1)}
	&\Re_\tinySEC^\tinyINT(t) > 3000
	\end{matrix}
	\right.
	\label{eqn:cond}
\end{equation}
\normalsize

\begin{equation}
	\psi\,= (0.790\,\ln(\Re_\tinySEC^\tinyINT(t)-1.64)^{-2}\ 
	\label{eqn:f}
\end{equation}

In case of laminar flow, the first equation of Equation Set \eqref{eqn:cond} is used. This is the Baehr correlation \cite{bergman2011fundamentals}, which computes the average \emph{Nusselt} number for forced convection in laminar flow inside a pipe, taking into consideration when both velocity and temperature profiles are not fully developed at the pipe inlet due to the viscous
forces in the laminar flow and thermal diffusivity. The longer the pipe, the closer is the configuration to a fully developed
velocity and temperature profile, and the \emph{Nusselt} number is to the limit value of 3.66. In order to take into consideration this entry region phenomenon, the \emph{Graetz} number is used, where the ratio between the internal radius and the length of the pipe is explicitly considered. This number is computed using Equation \eqref{eqn:Gra}.

\begin{equation}
	\Gz_\tinySEC^\tinyINT(t) = \dfrac{2 r_{\!\tinySEC}}{l_\tinySEC}\Pr_\tinySEC^\tinyINT\,\Re_\tinySEC^\tinyINT(t)
	\label{eqn:Gra}
\end{equation}

This correlation is valid for an uniform wall temperature. Considering that the fluid will not experience a high temperature increase/decrease and the assumption of an uniform temperature of the intermediate fluid, uniform wall temperature can be supposed.

In case of turbulent flow, the second equation of Equation Set \eqref{eqn:cond} is used. This is the Gnielinski correlation \cite{bergman2011fundamentals}, which uses the friction coefficient $\psi$. It has been chosen because its validity range matches that being used in the simulations.

The \emph{Reynolds} number for circular pipes and the \emph{Prandtl} number can be written as indicated in Equations \eqref{eqn:Re_int_sec} and \eqref{eqn:Pr_int_sec}.

\begin{equation}
	\Re_\tinySEC^\tinyINT(t) = \frac{2\,\dot m_\tinySEC(t)}{\mu_\tinySEC\,\pi\,r_{\!\tinySEC}\,n_\tinySEC} \
	\label{eqn:Re_int_sec}
\end{equation}

\begin{equation}
	\Pr_\tinySEC^\tinyINT = \frac{c_{\!p_{\,\tinySEC}}\mu_\tinySEC}{\kappa_\tinySEC}
	\label{eqn:Pr_int_sec}
\end{equation}

$\kappa_\tinySEC,\,c_{\!p_\tinySEC},\,$ and $\mu_\tinySEC$ are not constant, but computed as functions of the mass average temperature in the pipe, $\frac{1}{2}(T_\tinySEC^\tinyIN(t) + T_\tinySEC^\tinyOUT(t))$, and the pressure of the secondary fluid.

Regarding the refrigerant in the superheated vapour zone, Equations \eqref{eqn:alpha_int_refv} and \eqref{eqn:cond2} are used, where $\psi$ is the friction coefficient of turbulent fluid inside smooth pipes, computed using Equation \eqref{eqn:f}.

\begin{equation}
	\alpha_\tinyREFVAP^\tinyINT(t) = \frac{\kappa_\tinyREFVAP}{2\,r_{\!\tinyREF}}\,\Nu_\tinyREFVAP^\tinyINT(t)
	\label{eqn:alpha_int_refv}
\end{equation}

\scriptsize
\begin{equation}
	\left\{ 
	\begin{matrix} 
	\Nu_\tinySEC^\tinyINT(t) &=& \dfrac{\dfrac{3.66}{tanh[2.264\Gz_\tinyREFVAP^\tinyINT(t)^{-1/3}+1.7\Gz_\tinyREFVAP^\tinyINT(t)^{-2/3}]} \,+0.0499\Gz_\tinyREFVAP^\tinyINT(t)tanh(\Gz_\tinyREFVAP^\tinyINT(t)^{-1})}{tanh(2.432(\Pr_\tinyREFVAP^\tinyINT)^{1/6}\Gz_\tinyREFVAP^\tinyINT(t)^{-1/6})}\,\quad\qquad
	&\Re_\tinyREFVAP^\tinyINT(t) \le 3000
	\\ \\
	\Nu_\tinyREFVAP^\tinyINT(t) &=& \dfrac{\psi\!/\!8 (\Re_\tinyREFVAP^\tinyINT(t)-1000)
	\Pr_\tinyREFVAP^\tinyINT(t)}{1+12.7\,(\psi\!/\!8)^{1/2}(\Pr_\tinyREFVAP^\tinyINT(t)^{2/3}-1)}
	&\Re_\tinyREFVAP^\tinyINT(t) > 3000
	\end{matrix}
	\right.
	\label{eqn:cond2}
\end{equation}
\normalsize

Similarly, the \emph{Reynolds}, \emph{Prandtl}, and \emph{Graetz} numbers can be written as shown in Equations \eqref{eqn:Re_int_refv} -- \eqref{eqn:Gz_int_refv}.

\begin{equation}
	Re_\tinyREFVAP^\tinyINT(t) \,=\, \frac{2\,\dot m_\tinyREF(t)}{\mu_\tinyREFVAP\,\pi\,r_{\!\tinyREF}\,n_\tinyREF}
	\label{eqn:Re_int_refv}
\end{equation}

\begin{equation}
	\Pr_\tinyREFVAP^\tinyINT \,=\, \frac{c_{\!p_{\,\tinyREFVAP}}\mu_\tinyREFVAP}{\kappa_\tinyREFVAP}
	\label{eqn:Pr_int_refv}
\end{equation}

\begin{equation}
	\Gz_\tinyREFVAP^\tinyINT(t) = \dfrac{2 r_{\!\tinyREFVAP}}{(1-\zeta_\tinyREF(t))l_\tinyREF}\Pr_\tinyREFVAP^\tinyINT\,\Re_\tinyREFVAP^\tinyINT(t)
	\label{eqn:Gz_int_refv}
\end{equation}

$\kappa_\tinyREFVAP,\,c_{\!p_\tinyREFVAP},\,$ and $\mu_\tinyREFVAP$ can be computed as functions of the mass average temperature in the superheated vapour zone of the refrigerant pipe, $\frac{1}{2}(T_\tinyREF^\tinyLAT + T_\tinyREF^\tinyOUT(t))$, and the pressure of the refrigerant.

\subsection{Correlation via \emph{Klimenko's method}} \label{secAppendixInternalForcedConvectionPipesKlimenko}

This correlation is used to compute the internal heat transfer coefficient of the refrigerant in the two-phase zone.

In a first step, Klimenko's method infers whether the flow pattern is mainly bubbly or annular, by computing the following so-called $\phi-function$, computed using Equation \eqref{eqn:Kli1} \cite{KLIMENKO1988541,KLIMENKO19902073} .

\begin{equation}
	\phi_\tinyREFTWO(t) = \frac{G_\tinyREF(t) \,h_\tinyREF^\tinyLAT}{q_\tinyREFTWO(t)}
	\left[1 + \chi_\tinyREFTWO\, \left(\frac{\rho_\tinyREF^\tinyLATMIN}{\rho_\tinyREF^\tinyLATMAX}-1\right) \right]
	\left(\frac{\rho_\tinyREF^\tinyLATMAX}{\rho_\tinyREF^\tinyLATMIN}\right)^{\!\!1\!/3}
	\label{eqn:Kli1}
\end{equation}

$G_\tinyREF(t)$ is the refrigerant mass flux (rate of mass flow per unit area), while $q_\tinyREFTWO(t)$ is the internal heat flow (heat flow per unit area) in the two-phase zone, computed using Equations \eqref{eqn:Kli2} and \eqref{eqn:Kli3}.

\begin{equation}
	G_\tinyREF(t) = \frac{\dot m_\tinyREF(t)}{\pi\,r_\tinyREF^2\,n_\tinyREF}
	\label{eqn:Kli2}
\end{equation}

\begin{equation}
	q_\tinyREFTWO(t) \,=\, \frac{\dot Q_\tinyREFTWO(t)}{2\,\pi\,r_\tinyREF\,l_\tinyREF\,\zeta_\tinyREF(t)\,n_\tinyREF}
	\,=\, \frac{\dot m_\tinyREF(t)\,h_\tinyREF^\tinyLAT\,\left(\chi_\tinyREFTWO^\tinyOUT-\chi_\tinyREFTWO^\tinyIN\right)}{2\,\pi\,r_\tinyREF\,l_\tinyREF\,\zeta_\tinyREF(t)\,n_\tinyREF}
	\label{eqn:Kli3}
\end{equation}

The refrigerant input and output vapour qualities in the two-phase zone are, respectively, $\chi_\tinyREFTWO^\tinyIN=0.5549$ and $\chi_\tinyREFTWO^\tinyOUT=1.0$. The $\phi-function$ can be written as shown in Equation \eqref{eqn:Kli4}, where the mean vapour quality is used: $\chi_\tinyREFTWO = (\chi_\tinyREFTWO^\tinyIN +\chi_\tinyREFTWO^\tinyOUT)/2 = 0.7775$.

\begin{equation}
	\phi_\tinyREFTWO(t) = \frac{2\,l_\tinyREF\,\zeta_\tinyREF(t)}{r_\tinyREF\,\left(\chi_\tinyREFTWO^\tinyOUT-\chi_\tinyREFTWO^\tinyIN\right)} \left[1 + \chi_\tinyREFTWO\, \left(\frac{\rho_\tinyREF^\tinyLATMIN}{\rho_\tinyREF^\tinyLATMAX}-1\right) \right]\left(\frac{\rho_\tinyREF^\tinyLATMAX}{\rho_\tinyREF^\tinyLATMIN}\right)^{\!\!1\!/3}
	\label{eqn:Kli4}
\end{equation}

According to Klimenko's method, three intervals are stated for the computation of the \emph{Nusselt} number, depending on the value of the $\phi-function$, as indicated in Equation Set \eqref{eqn:Kli5}, whose intermediate parameters are computed with Equations \eqref{eqn:Kli6} -- \eqref{eqn:Kli10}. Note that $l_\tinyREFTWO'$ is the relevant characteristic length, being computed using Equation \eqref{eqn:Kli8}.

\footnotesize
\begin{equation}
	\left\{ 
	\begin{matrix}
	\Nu_\tinyREFTWO^\tinyINT(t) &=& 0.0076\,\,\mathsf{q}_\tinyREFTWO^{0.6}\,\mathsf{p}_\tinyREFTWO^{0.5} \left(\Pr_\tinyREFTWO^\tinyINTLATMIN\right)^{-1\!/3}
	\left(\frac{\kappa_\tinyREF^\tinyWALL}{\kappa_\tinyREF^\tinyLATMIN}\right)^{0.15}\,\quad\qquad
	& \phi_\tinyREFTWO(t) \le 12000 \\ \\
	\Nu_\tinyREFTWO^\tinyINT(t) &=& 0.087 \,\left(\Re_\tinyREFTWO^{\tinyINT'}\right)^{0.6}\,\left(\Pr_\tinyREFTWO^\tinyINTLATMIN\right)^{1/6}\left(\frac{\rho_\tinyREF^\tinyLATMAX}{\rho_\tinyREF^\tinyLATMIN}\right)^{0.2} \left(\frac{\kappa_\tinyREF^\tinyWALL}{\kappa_\tinyREF^\tinyLATMIN}\right)^{0.09}\,\quad\qquad
	&\phi_\tinyREFTWO(t) > 20000 \\ \\
	\Nu_\tinyREFTWO^\tinyINT(t) &=& \text{Maximum of the previous ones}\,\quad\qquad &
	12000 < \phi_\tinyREFTWO(t) \le 20000
	\end{matrix}
	\right.
	\label{eqn:Kli5}
\end{equation}
\normalsize

\begin{equation}
	\mathsf{q}_\tinyREFTWO(t) = \frac{q_\tinyREFTWO(t)\,l_\tinyREFTWO'}{h_\tinyREF^\tinyLAT\,\frac{\rho_\tinyREF^\tinyLATMAX}{\rho_\tinyREF^\tinyLATMIN}\frac{\kappa_\tinyREF^\tinyLATMIN}{c_{\!p_{\,\tinyREF}}^\tinyLATMIN}}
	\label{eqn:Kli6}
\end{equation}

\begin{equation}
	\mathsf{p}_\tinyREFTWO(t) = \frac{P_\tinyREF\,l_\tinyREFTWO'}{\sigma_\tinyREF^\tinyLATMIN}
	\label{eqn:Kli7}
\end{equation}

\begin{equation}
	l_\tinyREFTWO'  \,=\, \left(\frac{\sigma_\tinyREF^\tinyLATMIN}{g\left(\rho_\tinyREF^\tinyLATMIN \!-\! \rho_\tinyREF^\tinyLATMAX\right)}\right)^{1/2}
	\label{eqn:Kli8}
\end{equation}

\begin{equation}
	\Re_\tinyREFTWO^{\tinyINT'}(t) \,=\, \frac{\dot m_\tinyREF(t)\,l_\tinyREFTWO'}{\mu_\tinyREF^\tinyLATMIN\,\pi\,r_{\!\tinyREF}^2\,n_\tinyREF} \left[1 + \chi_\tinyREFTWO\, \left(\frac{\rho_\tinyREF^\tinyLATMIN}{\rho_\tinyREF^\tinyLATMAX}-1\right) \right]
	\label{eqn:Kli9}
\end{equation}

\begin{equation}
	\Pr_\tinyREFTWO^\tinyINTLATMIN \,=\, \frac{c_{\!p_{\,\tinyREF}}^\tinyLATMIN \mu_\tinyREF^\tinyLATMIN}{\kappa_\tinyREF^\tinyLATMIN}
	\label{eqn:Kli10}
\end{equation}

The internal convective coefficient of the refrigerant in the two-phase zone can be obtained using Equation \eqref{eqn:Kli11}.

\begin{equation}
	\alpha_\tinyREFTWO^{\tinyINT'}(t) \,=\, \frac{\kappa_\tinyREF^\tinyLATMIN}{l_\tinyREFTWO'}\,\Nu_\tinyREFTWO^\tinyINT(t)
	\label{eqn:Kli11}
\end{equation}

Klimenko suggests, however, to ''average'' the previous estimation by using the internal forced convective coefficient of the refrigerant in hypothetical liquid phase with zero vapour quality, computed using Equation \eqref{eqn:Kli12}.

\begin{equation}
	\alpha_\tinyREFTWO^\tinyINTLATMIN(t) \,=\, \frac{\kappa_\tinyREF^\tinyLATMIN}{2\,r_{\!\tinyREF}}\,\Nu_\tinyREFTWO^\tinyINTLATMIN(t)
	\label{eqn:Kli12}
\end{equation}

The particular correlation depends on the instantaneous value of the \emph{Reynolds} number, as indicated in Equation Set \eqref{eqn:Kli13},  where $\psi$ is the friction coefficient of turbulent fluid inside smooth pipes, computed using Equation \eqref{eqn:f}. Moreover, the corresponding \emph{Reynolds} number, referred to the liquid phase, is computed as shown in Equation \eqref{eqn:Kli14}.

\scriptsize
\begin{equation}
	\left\{ 
	\begin{matrix} 
	\Nu_\tinyREFTWO^\tinyINTLATMIN(t) &=& \dfrac{\dfrac{3.66}{tanh[2.264\Gz_\tinyREFTWO^\tinyINTLATMIN(t)^{-1/3}+1.7\Gz_\tinyREFTWO^\tinyINTLATMIN(t)^{-2/3}]} \,+0.0499\Gz_\tinyREFTWO^\tinyINTLATMIN(t)tanh(\Gz_\tinyREFTWO^\tinyINTLATMIN(t)^{-1})}{tanh(2.432(\Pr_\tinyREFTWO^\tinyINTLATMIN)^{1/6}\Gz_\tinyREFTWO^\tinyINTLATMIN(t)^{-1/6})}\,\quad\qquad
	&\Re_\tinyREFTWO^\tinyINTLATMIN(t) \le 3000
	\\ \\
	\Nu_\tinyREFTWO^\tinyINTLATMIN(t) &=&
	\dfrac{\psi\!/\!8 (\Re_\tinyREFTWO^\tinyINTLATMIN(t)-1000)
	\Pr_\tinyREFTWO^\tinyINTLATMIN(t)}{1+12.7\,(\psi\!/\!8)^{1/2}(\Pr_\tinyREFTWO^\tinyINTLATMIN(t)^{2/3}-1)}
	&\Re_\tinyREFTWO^\tinyINTLATMIN(t)
	> 3000
	\end{matrix}
	\right.
	\label{eqn:Kli13}
\end{equation}
\normalsize

\begin{equation}
	\Re_\tinyREFTWO^\tinyINTLATMIN(t) \,=\, \frac{2\,\dot
	m_\tinyREF(t)}{\mu_\tinyREF^\tinyLATMIN\,\pi\,r_{\!\tinyREF}\,n_\tinyREF}\,\left(1-\chi_\tinyREFTWO\right)
	\label{eqn:Kli14}
\end{equation}

Note that $\rho_\tinyREF^\tinyLATMAX,\,\rho_\tinyREF^\tinyLATMIN,\,\kappa_\tinyREF^\tinyLATMIN,\,c_{\!p_\tinyREF}^\tinyLATMIN,\,\mu_\tinyREF^\tinyLATMIN\,$, and $\sigma_\tinyREF^\tinyLATMIN$ are only functions of the refrigerant pressure. Finally, the convective coefficient is computed as indicated in Equation \eqref{eqn:Kli15}.

\begin{equation}
	\alpha_\tinyREFTWO^{\tinyINT}(t) \,=\, \left( \alpha_\tinyREFTWO^{\tinyINT'}(t)^3 \,+\, \alpha_\tinyREFTWO^{\tinyINTLATMIN}(t)^3 \right)^{1/3}
	\label{eqn:Kli15}
\end{equation}


\section{Summary of equations of the continuous model} \label{appendixEquationSummary}

This Appendix summarises the input variables and unknown variables included in the continuous model, as well as the nonlinear equation set to be solved.

In our continuous model, we have the set of known input variables (not to mention other steady variables, such as working pressures) indicated in Equation \eqref{apB_knownset}.

\begin{equation}
	\{ T_\tinyREF^\tinyIN(t),\, T_\tinySEC^\tinyIN(t),\, T_\tinyENV(t),\, \dot m_\tinyREF(t),\, \dot m_\tinySEC(t),\, h_\tinyREF^\tinyIN(t) \}
	\label{apB_knownset}
\end{equation}

The set of eighteen unknowns shown in Equation \eqref{apB_unknownset} must be solved.

\begin{equation}
	\begin{aligned}
	\{ &T_\tinyINT(t),\, T_\tinyPCM(t),\, T_\tinyPCM^\tinyWALL(t),\, T_\tinyREF^\tinyOUT(t),\, h_\tinyREF^\tinyOUT(t),\, T_\tinyREFTWO^\tinyWALL(t),\, T_\tinyREFVAP^\tinyWALL(t),\,T_\tinySEC^\tinyOUT(t),\,T_\tinySEC^\tinyWALL(t),\, ... \\
	... \,&r(t),\,r_\tinyPCM(t),\,\zeta_\tinyREF(t),\,\dot Q_\tinyPCM(t),\,\dot Q_\tinyREF(t),\,\dot Q_\tinyREFTWO(t),\,\dot Q_\tinyREFVAP(t),\,\dot Q_\tinySEC(t),\, \dot Q_\tinyENV(t) \} \\
	\end{aligned}
	\label{apB_unknownset}
\end{equation}

To compute the values of the eighteen unknown variables, the eighteen equations indicated in Equation Set \eqref{apB_eqset} are imposed.

\begin{subequations}
	
	\begin{equation}
		-m_\tinyINT\,c_{\!p_\tinyINT} \dot T_\tinyINT(t) \,=\, n_\tinyREF\, \dot Q_\tinyREF(t) + n_\tinySEC\, \dot Q_\tinySEC(t) + n_\tinyPCM \, \dot Q_\tinyPCM(t) + \dot Q_\tinyENV(t)
		\label{apB_eqset1}
	\end{equation}
	
	\begin{equation}
		\dot Q_\tinyPCM(t) \,=\, \frac{T_\tinyPCM^\tinyWALL(t) - T_\tinyPCM(t)}{R_\tinyPCM^\tinyCONDINT(t) + R_\tinyPCM^\tinyCONDWALL}
		\label{apB_eqset2}
	\end{equation}
	
	\begin{equation}
		\dot Q_\tinyPCM(t) \,=\, \frac{T_\tinyINT(t) - T_\tinyPCM(t)}{R_\tinyPCM^\tinyCONDINT(t) + R_\tinyPCM^\tinyCONDWALL + R_\tinyPCM^\tinyCONVEXT(t)}
		\label{apB_eqset3}
	\end{equation}
	
	\begin{equation}
		\left\{ 
		\begin{matrix}
			\dot Q_\tinyPCM(t) \,&=\, \rho_\tinyPCM ^\tinyLATMAX \, h_\tinyPCM^\tinyLAT 4\pi\, r(t)^2 \,\dot r(t) \quad\qquad
			& \text{Charge} \\ \\
			\dot Q_\tinyPCM(t) \,&=\, - \rho_\tinyPCM ^\tinyLATMAX \, h_\tinyPCM^\tinyLAT 4\pi\, r(t)^2 \,\dot r(t) \quad\qquad
			& \text{Discharge} \\ \\
		\end{matrix}
		\right.
		\label{apB_eqset4}
	\end{equation}
	
	\begin{equation}
		T_\tinyPCM(t) \,\approx\, T_\tinyPCM^\tinyLAT \qquad \text{During charging and discharging operations}
		\label{apB_eqset5}
	\end{equation}
	
	\begin{equation}
		\left\{ 
		\begin{matrix}
		\dot r_\tinyPCM(t) \,=\, \dfrac{\rho_\tinyPCM ^\tinyLATMIN \,-\, \rho_\tinyPCM ^\tinyLATMAX}{\rho_\tinyPCM ^\tinyLATMIN} \, \dfrac{{r(t)}^2}{{r_\tinyPCM(t)}^2} \, \dot r(t) \, \,<\, 0  \quad\qquad
		& \text{Charge} \\ \\
		\dot r_\tinyPCM(t) \,=\, \,-\, \dfrac{\rho_\tinyPCM ^\tinyLATMIN \,-\, \rho_\tinyPCM ^\tinyLATMAX}{\rho_\tinyPCM ^\tinyLATMAX} \, \dfrac{{r(t)}^2}{{r_\tinyPCM(t)}^2} \, \dot r(t) \, \,>\, 0 \quad\qquad
		& \text{Discharge} \\ \\
		\end{matrix}
		\right.
		\label{apB_eqset6}
	\end{equation}
	
	\begin{equation}
		\dot Q_\tinySEC(t) \,=\, C_\tinySEC(t)\,(T_\tinySEC^\tinyOUT(t) - T_\tinySEC^\tinyIN(t))
		\label{apB_eqset7}
	\end{equation}
	
	\begin{equation}
		\dot Q_\tinySEC(t) \,=\, \frac{T_\tinySEC^\tinyWALL(t) - \frac{1}{2}\left(T_\tinySEC^\tinyIN(t) + T_\tinySEC^\tinyOUT(t)\right)}{R_\tinySEC^\tinyCONVINT(t) + R_\tinySEC^\tinyCONDWALL}
		\label{apB_eqset8}
	\end{equation}
	
	\begin{equation}
		\dot Q_\tinySEC(t) \,=\, \varepsilon_\tinySEC(t)\, C_\tinySEC(t)\,(T_\tinyINT(t) - T_\tinySEC^\tinyIN(t))
		\label{apB_eqset9}
	\end{equation}
	
	\begin{equation}
		\dot Q_\tinyENV(t) \,\approx\, 0 
		\label{apB_eqset10}
	\end{equation}
	
	\begin{equation}
		\dot Q_\tinyREF(t) \,=\, \dot Q_\tinyREFTWO(t)\,+\, \dot Q_\tinyREFVAP(t)
		\label{apB_eqset11}
	\end{equation}

	\begin{equation}
		\text{\emph{Mode} 1}
		\left\{ 
		\begin{matrix}
		\dot Q_\tinyREFTWO(t) \,=\, \dot m_\tinyREF\!(t)\, (h_\tinyREF^\tinyLATMAX - h_\tinyREF^\tinyIN(t)) \\ \\
		\dot Q_\tinyREFTWO(t) \,=\, \frac{T_\tinyREFTWO^\tinyWALL(t)-T_\tinyREF^\tinyLAT}{R_\tinyREFTWO^\tinyCOND(t) + R_\tinyREFTWO^\tinyCONVINT(t)} \\ \\
		\dot Q_\tinyREFTWO(t) \,=\, \frac{T_\tinyINT(t) - T_\tinyREF^\tinyIN(t)}{R_\tinyREFTWO^\tinyCONVINT\!(t)  + R_\tinyREFTWO^\tinyCONDWALL\!(t)  + R_\tinyREFTWO^\tinyCONVEXT\!(t)} \\ \\
		\dot Q_\tinyREFVAP(t) \,=\, C_\tinyREFVAP(t)\,(T_\tinyREF^\tinyOUT(t) - T_\tinyREF^\tinyLATMAX) \\ \\
		\dot Q_\tinyREFVAP(t) \,=\, \frac{T_\tinyREFVAP^\tinyWALL(t) - \frac{1}{2}\left(T_\tinyREF^\tinyLATMAX + T_\tinyREF^\tinyOUT(t)\right)}{R_\tinyREFVAP^\tinyCONVINT(t) + R_\tinyREFVAP^\tinyCONDWALL} \\ \\
		\dot Q_\tinyREFVAP(t) \,=\, \varepsilon_\tinyREFVAP(t)\, C_\tinyREFVAP(t)\,(T_\tinyINT(t) - T_\tinyREF^\tinyIN(t)) \\ \\
		\dot Q_\tinyREFVAP(t) \,=\, \dot m_\tinyREF\!(t)\, (h_\tinyREF^\tinyOUT(t) - h_\tinyREF^\tinyLATMAX) \\ \\
		\end{matrix}
		\right.
		\label{apB_eqset12a}
	\end{equation}

	\begin{equation}
		\text{\emph{Mode} 2}
		\left\{ 
		\begin{matrix}
		\zeta_\tinyREF(t) \,=\, 1 \\ \\
		\dot Q_\tinyREFTWO(t) \,=\, \dot m_\tinyREF\!(t)\, (h_\tinyREF^\tinyOUT(t) - h_\tinyREF^\tinyIN(t)) \\ \\
		\dot Q_\tinyREFTWO(t) \,=\, \frac{T_\tinyREFTWO^\tinyWALL(t)-T_\tinyREF^\tinyLAT}{R_\tinyREFTWO^\tinyCOND(t) + R_\tinyREFTWO^\tinyCONVINT(t)} \\ \\
		\dot Q_\tinyREFTWO(t) \,=\, \frac{T_\tinyINT(t) - T_\tinyREF^\tinyIN(t)}{R_\tinyREFTWO^\tinyCONVINT\!(t)  + R_\tinyREFTWO^\tinyCONDWALL\!(t)  + R_\tinyREFTWO^\tinyCONVEXT\!(t)} \\ \\
		\dot Q_\tinyREFVAP(t) \,=\, 0 \\ \\
		T_\tinyREF^\tinyOUT(t) \,=\, T_\tinyREF^\tinyLAT \\ \\
		T_\tinyREFVAP^\tinyWALL \quad \text{is not defined in \emph{mode} 2} \\ \\
		\end{matrix}
		\right.
		\label{apB_eqset12b}
	\end{equation}
	
	\label{apB_eqset}
	
\end{subequations}


\section*{References}

\bibliographystyle{IEEEtran}
\bibliography{bibliografia}

\end{document}